\documentclass[prb,twocolumn,superscriptaddress]{revtex4}

\usepackage{amsmath}
\usepackage{graphicx}
\usepackage{amssymb}
\usepackage{amsthm, amscd}
\usepackage{amsfonts}
\usepackage{dsfont}
\usepackage{cancel}
\usepackage{times}
\usepackage{pstricks}
\usepackage{wasysym}
\usepackage{ulem}
\usepackage[dvips]{epsfig}
\begin{document}

%%%%%%%%%%%%

\title{Boundary Conformal Field Theory and Tunneling of Edge
  Quasiparticles\\ 
in non-Abelian Topological States}

%\title{Resonant tunneling of edge quasiparticles in the
%$\nu=5/2$ non-Abelian quantum Hall state}

\author{Paul Fendley}
\affiliation{Department of Physics, University of Virginia,
Charlottesville, VA 22904-4714}
\author{Matthew P.A. Fisher}
\affiliation{Microsoft Research, Station Q, CNSI Building,
University of California, Santa Barbara, California 93106, USA}
\affiliation{Department of Physics, University of California,
Santa Barbara, California, 93106, USA}
\author{Chetan Nayak}
\affiliation{Microsoft Research, Station Q, CNSI Building,
University of California, Santa Barbara, California 93106, USA}
\affiliation{Department of Physics, University of California,
Santa Barbara, California, 93106, USA}

\date{February 2, 2009}

\begin{abstract}
We explain how (perturbed) boundary conformal field theory allows us
to understand the tunneling of edge quasiparticles in non-Abelian
topological states. The coupling between a bulk non-Abelian
quasiparticle and the edge is due to resonant tunneling
to a zero mode on the quasiparticle, which causes the
zero mode to hybridize with the edge. This can be reformulated
as the flow from one conformally-invariant boundary
condition to another in an associated critical statistical
mechanical model. Tunneling from one edge to another
at a point contact can split the system in two,
either partially or completely. This can be reformulated
in the critical statistical mechanical model as the
flow from one type of defect line to another.
We illustrate these two phenomena in detail
in the context of the $\nu=5/2$ quantum Hall state and the
critical Ising model. We briefly discuss the case
of Fibonacci anyons and conclude by explaining
the general formulation and its physical interpretation. 
\end{abstract}

\maketitle

%%%%%%%%%%%%

\section{Introduction}

The remarkable features of non-Abelian topological phases, including
their potential use for quantum computation
\cite{Kitaev97,Freedman02a,Nayak08}, stem from
the non-integer number of internal degrees of
freedom per quasiparticle. Namely, the number
of states in the $N$-quasiparticle Hilbert spaces for $N$ large grows
as $d^N$, where $d$ is the {\it quantum dimension}.  For instance, the
most promising models of the $\nu=5/2$ quantum Hall state
\cite{Moore91,Greiter92,Nayak96c,Lee07,Levin07} have charge-$e/4$
quasiparticles with $d=\sqrt{2}$ as do flux $hc/2e$ vortices in a
$p+ip$ superconductor
\cite{Read00,Ivanov01,Stern04,Stone06}.  Models
supporting universal quantum computation, including one which may be
relevant to the $\nu=12/5$ quantum Hall state \cite{Read99,Bishara08},
have quasiparticles with $d=(1+\sqrt{5})/2$, the golden ratio.  In
chiral topological phases, there are necessarily gapless excitations
at the edge of the system.  When a bulk quasiparticle is close to the
edge, the degeneracy is lifted because of its interactions with these
gapless excitations. In this paper, we uncover the dynamics by which
the degeneracy of internal degrees of freedom is lifted.

A useful tool in our analysis is to exploit the equivalence of a
quantum system in $d$ spatial dimensions and a classical system in
$d+1$ spatial dimensions. The bulk-edge dynamics in a topological
state then can be described by the flow between different
conformally-invariant boundary conditions in an associated critical 2D
statistical mechanical model. For example, the coupling to the edge of
a charge-$e/4$ quasiparticle at $\nu=5/2$ (or of a flux $hc/2e$ vortex
in a $p+ip$ superconductor) is equivalent to the imposition of a
magnetic field at the boundary of the critical 2D Ising model on the
half-plane, causing a flow from free to fixed boundary conditions.

Backscattering between edges of a topological state at a point contact
can also be understood as a flow between conformally-invariant
boundary conditions. Namely, by ``squashing'' the edge of the system
onto a line segment, which is then folded about the point contact (see
Fig.\ \ref{fig:folded-drop} below), inter-edge backscattering can be
understood in terms of two copies of the
associated critical 2D statistical mechanical model coupled only at
their boundary. Such a rephrasing allows us to place our earlier work
\cite{Fendley06,Fendley07a,Fendley07b} on charge-$e/4$ quasiparticle
backscattering at $\nu=5/2$ in a wider context.  The squashing
procedure also allows us to study more complicated topologies such as
the annulus and situations in which there are multiple bulk
quasiparticles.

In this paper, we focus mainly on the case of a chiral Majorana
fermion edge mode, which is the edge theory of a $p+ip$ superconductor
and is the neutral sector of the edge theory of the $\nu=5/2$ quantum
Hall state \cite{Milovanovic96}. (Some of our results do not apply
to a $p+ip$ superconductor because its vortices are
essentially classical as far as their motion is concerned,
but they do apply  to a topological state which
may be viewed as a quantum-disordered $p+ip$
superconductor resulting from the condensation of
$hc/e$ vortices. We will simply use Majorana fermion edge mode
to refer to edge of this state and the neutral sector of the
edge theory of the proposed non-Abelian $\nu=5/2$ quantum Hall
states \cite{Moore91,Lee07,Levin07}.)
The classical analog is the critical
Ising model, whose boundary conditions \cite{Chatterjee94} and defect
lines \cite{Oshikawa97} have been analyzed in depth. We make a
connection with these results, leading to a simple interpretation for
a critical line of defect boundary conditions (`continuous Neumann')
which has no simple interpretation in Ising language.  
We will discuss briefly the added complications arising from the
presence of a charged mode in the $\nu=5/2$ state.  We will also
mention how our results can be generalized to other conformal field
theories, and will briefly discuss the case of the $k=3$ Read-Rezayi
state and $\mathbb{Z}_3$ parafermions.

In section \ref{sec:Majorana-Ising}, we discuss the mapping between
the edge theory of (the neutral sector of) the $5/2$ quantum Hall
state and the critical $2D$ Ising field theory.  In section
\ref{sec:free-fixed-flow}, we analyze the effect of Majorana fermion
tunneling between a bulk vortex and the edge, showing that this is the
same problem as a magnetic field applied to the boundary of a critical
Ising model \cite{Chatterjee94}. In section \ref{sec:point-contact},
we show how the effects of a point contact can be expressed in terms
of a boundary problem by folding the system. The folding us allows us
to bosonize, and to utilize the results of Oshikawa and Affleck for
the Ising model with a defect line \cite{Oshikawa97}.  In section
\ref{sec:neutral-pt-contact}, we analyze the effect of interedge
backscattering at a point contact in a Majorana fermion edge mode, and
discuss in depth two critical lines of boundary fixed points. We also
discuss the closely related case of the Moore-Read Pfaffian state.  In
section \ref{sec:flows} we analyze flows between these fixed points,
and study the entropy drops. We extend our analysis to allow for an
arbitrary number of quasiparticles in section \ref{sec:many-qps}. In
section \ref{sec:potts}, we generalize our results to a different
topological state, supporting Fibonacci anyons.  Finally, in section
\ref{sec:discussion}, we discuss our results in the larger context of
topological phases and the transitions between them.

\section{Mapping a Majorana fermion edge mode
to the critical $2D$ Ising model on a strip}
\label{sec:Majorana-Ising}

Consider a very large quantum Hall droplet at filling $\nu=5/2$ which
we assume initially is in a Moore-Read Pfaffian state \cite{Moore91}.
Circumnavigating the droplet are gapless chiral edge modes
\cite{Wen91}: a bosonic charge mode, $\phi_c$, and a neutral Majorana
fermion, $\psi$. \cite{Milovanovic96} Initially we focus our attention
exclusively on the neutral sector -- a chiral Majorana fermion --
which is formally equivalent to the edge of a $p+ip$ superconductor
 \cite{Read00,Ivanov01}.

Taking the circumference of the droplet to be $2L$ it is convenient to
``squash'' this chiral system into an effectively one-dimensional
model with both left and right movers.  Reformulating the problem on a
strip allows us to treat tunneling as a problem in 1+1-dimensional
boundary field theory, or equivalently, a quantum impurity
problem. There is an enormous literature on such problems, much of it
following the seminal paper \cite{Cardy89}.  For the case of a single
Majorana fermion, many results which are useful for us have already been
obtained in this context, in particular
Refs.\ \onlinecite{Chatterjee94,Oshikawa97}.  Thus, we introduce right-
and left- moving fields, $\psi_R(x)$ and $\psi_L(x)$, which are
functions of an $x-$coordinate lying in the interval $[0,L]$.  The
action describing the edge dynamics is
\begin{equation}
\label{eqn:strip-action}
S_0= \int dt  \int_{0}^{L} dx \,{\cal L}_0,
\end{equation}
with Lagrangian density,
\begin{equation}
\label{eqn:fermion-Lagrangian}
{\cal L}_0 = i \psi_R({\partial_t}+{v_n}\partial_x)\psi_R +
 i \psi_L({\partial_t}-{v_n}\partial_x)\psi_L .
\end{equation}
The edge modes have dispersion $\epsilon(k) = v_n k$ with the momenta
$k \ge 0$ chosen to satisfy the appropriate boundary conditions which
we shall discuss momentarily.  Thus, the neutral sector of the edge of
the Moore-Read Pfaffian state or, equivalently, the edge of a $p+ip$
superconductor is simply given by a non-chiral gapless Majorana
fermion on the strip $x\in[0,L]$, $\tau\in[-\infty,\infty]$.  (At
non-zero temperature, the length in the Euclidean time direction
$\tau$ is also finite.)

In order to complete the mapping to the strip,
we must specify the boundary conditions at the
two ends of the strip, $x=0,L$.
These are independent of the exact shape of the droplet,
since the edge theory is conformally-invariant.
They are, instead, determined by how the fermionic field behaves
as one makes a circuit of the droplet. For the $p+ip$
superconductor (or Moore-Read Pfaffian state),
this depends on the number of $hc/2e$
vortices (or charge-$e/4$ quasiparticles) in the
bulk. When there are no vortices (or an even number),
the edge fermion behaves as fermions typically do
under rotations of $2\pi$: the fermionic field picks up a minus sign,
so that it is {\em antiperiodic}. (This may be seen explicitly
from the Bogoliubov-de Gennes equations in the $p+ip$
superconducting case or from the lowest Landau level wavefunctions
in the Moore-Read Pfaffian case.) The presence of a single
vortex (or an odd number) flips this to {\em periodic};
the vortex can be viewed as introducing a branch
cut.  An even number of vortices therefore leaves the fermionic field
antiperiodic, while an odd number makes it periodic.

Once we have squashed to the strip, we need
boundary conditions at the ends of the line segment which
reflect left movers into right movers at $x=0$
and right movers back into left movers at
$x=L$. Instead of imposing these boundary
conditions by hand, it is much more
convenient instead to add boundary terms $L_b$
to the action so that the boundary conditions are
consequences of the equations of motion.
This means that the original boundary conditions
are treated on an equal footing as those induced by tunneling,
making it much simpler to understand the flows between
different boundary conditions. 
With this idea in mind, we modify the action to
\begin{equation}
\label{eqn:action-with-b}
S = {S_0} + \int dt \, {L_b}
\end{equation}
where the boundary terms in the absence of tunneling are
\begin{equation}
{L_b} = - ia v_n\psi_L(0)\psi_R(0) +  ibv_n\psi_L(L)\psi_R(L)\ .
\label{bterm}
\end{equation}
For Grassman variables $\alpha,\beta$, we adopt the complex
conjugation convention $(\alpha \beta)^* = \beta^*\alpha^*$, and for
Majorana fermions we have $\psi_L^*=\psi_L$ and $\psi_R^*=\psi_R$. The
term $L_b$ then indeed obeys $(L_b)^*=L_b$ if $a$ and $b$ are real.
Once $L_b$ is included, the equations of motion for $\psi_L$ and
$\psi_R$ are found by varying them independently. Varying $\psi_R$
in (\ref{eqn:action-with-b}) yields
\begin{equation}
\psi_R(0)=a\,\psi_L(0),\qquad
\psi_R(L)=b\,\psi_L(L)\ 
\label{bcfermion}
\end{equation} 
for the equations of motion at the boundaries; the terms on the
left-hand sides result from a surface contribution from
$S_0$.
Varying $\psi_L$ gives the same equations with $L$ and $R$
reversed, so consistency demands that $a^2=1$ and $b^2=1$.

The values of $a$ and $b$ in the boundary conditions depend on the
number of vortices in the bulk. In the unsquashed geometry, the
boundary conditions are antiperiodic when there is an even number
$N_v$ of vortices in the bulk. This means that we must have
$ab=-1$ to reproduce this in the squashed geometry.
It turns out to be more convenient, and to agree with the
natural choice in conformal field theory \cite{Chatterjee94},
to define $\psi_L$ and $\psi_R$ so that $a=-1$ and $b=1$
when $N_v=0$. We discuss the reasons for this below.

The introduction of bulk quasiparticles seems to complicate matters
considerably, since the boundary conditions for
the edge fermion depend on whether $N_v$ is even or odd.
However, a main point of one of our earlier
papers is that key topological properties of the bulk quasiparticles
can be taken into account by understanding edge properties
\cite{Fendley07b}. In this paper we show that, equivalently, such
effects can be incorporated via the boundary conditions. 

We discuss this in depth below, but let us begin here with the
simplest situation $e/4$-quasiparticle
pinned in the bulk of the sample.  The boundary conditions on the
chiral fermion in the original geometry are now periodic, because the
vortex introduces a branch cut. This branch cut is very important:
with it, the fermion has a zero mode on the edge, i.e.\ a solution to
the equations of motion having zero energy \cite{Ginsparg}. In the
squashed picture, it is convenient to make the branch cut go through
one of the points which becomes a boundary after squashing, so
including the cut amounts to modifying the boundary conditions to
$a=b$. The choice of whether $a=b=1$ or $a=b=-1$ is equivalent to
having the branch cut go through the left or right.

It is very useful to rephrase the preceding in
the language of the Ising model.
The Ising model can be described by a single
non-chiral Majorana fermion: the post-squashing Lagrangian
density (\ref{eqn:fermion-Lagrangian}) is precisely that of the
transverse field Ising model at its quantum critical
point on a line segment.
If we continue to Euclidean time, this corresponds to 
classical Ising model at its (bulk) critical point on a strip.
Equivalently, it corresponds to the $1D$ quantum
transverse field Ising model on a finite chain at its
critical point. 
There are two possible boundary conditions which
preserve scale invariance, known as
``free'' and ``fixed'' in terms of Ising spins. The free boundary
condition corresponds to having the end spin in the
quantum transverse field Ising chain
unconstrained, while the fixed boundary condition corresponds
to fixing that spin to be a particular value for all time
(or, in the classical $2D$ Ising model, to fixing all of the
spins along the boundary). 

%``free'' and ``fixed'' in terms of Ising spins. Free boundary
%conditions corresponds to having the end spin in the chain
%unconstrained, while fixed boundary conditions correspond to fixing
%that spin to be a particular value for all time. 

Relating the fermionic boundary conditions (\ref{bcfermion}) to Ising
ones precisely is a little subtle. Having no bulk vortices ($-a=b=1$)
corresponds to having fixed boundary conditions at {\em both} $x=0$
and $x=L$, even though $a=-b$ naively seems to imply that the boundary
conditions are different at the two ends. This can be seen indirectly
by comparing the partition functions for the boundary system
\cite{Cardy86b,Cardy89} with those for the topological state
\cite{Fendley07b}. A direct proof \cite{Chatterjee94} follows by first
considering Euclidean spacetime to be the upper-half-plane. Then one
has $\psi_L=-\psi_R$ along the $x$-axis for fixed boundary conditions,
and $\psi_L=\psi_R$ for free boundary conditions.  To find the
corresponding boundary conditions on the strip $x\in[0,L]$, we need to
conformally transform it to the upper-half-plane. We map
$z=e^{\tau+i\sigma}$, $\overline{z} =e^{\tau-i\sigma}$, which takes
the strip $\sigma=\pi x/L\in[0,\pi]$, $\tau\in[-\infty,\infty]$ to the
upper half-plane $\text{Im}\, z \geq 0$. Right-moving fields are then
functions of $z$, while left movers depend on $\overline{z}$.  Under
this conformal transformation, a dimension $D$ function of $z$
transforms along the boundary $\hbox{Im}\,z=0$ as $f(z)\to
\left(\frac{\partial z}{\partial\tau}\right)^D f(\tau+i\sigma)$, and
likewise $\overline{f}(\overline{z}) \to
\left(\frac{\partial\overline{z}}{\partial\tau}\right)^{\overline{D}}
f(\tau-i\sigma)$. This means that for $\sigma=0$, this conformal
transformation leaves the boundary condition unchanged, i.e. $a=-1$
for fixed and $a=1$ for free.  However, for $x=L$ (i.e.\
$\sigma=\pi$), the conformal transformation yields a factor $e^{i\pi
D}$ for the right movers and $e^{-i\pi \overline{D}}$ for the left
movers. Since the fermions have dimension $1/2$, the fixed boundary
condition here is modified to $\psi_L(L)=\psi_R(L)$ while free is now
$\psi_L(L)=-\psi_R(L)$, so that $b=1$ for fixed and $b=-1$ for free.

\begin{figure}[h]
\centerline{\includegraphics[width=85mm]{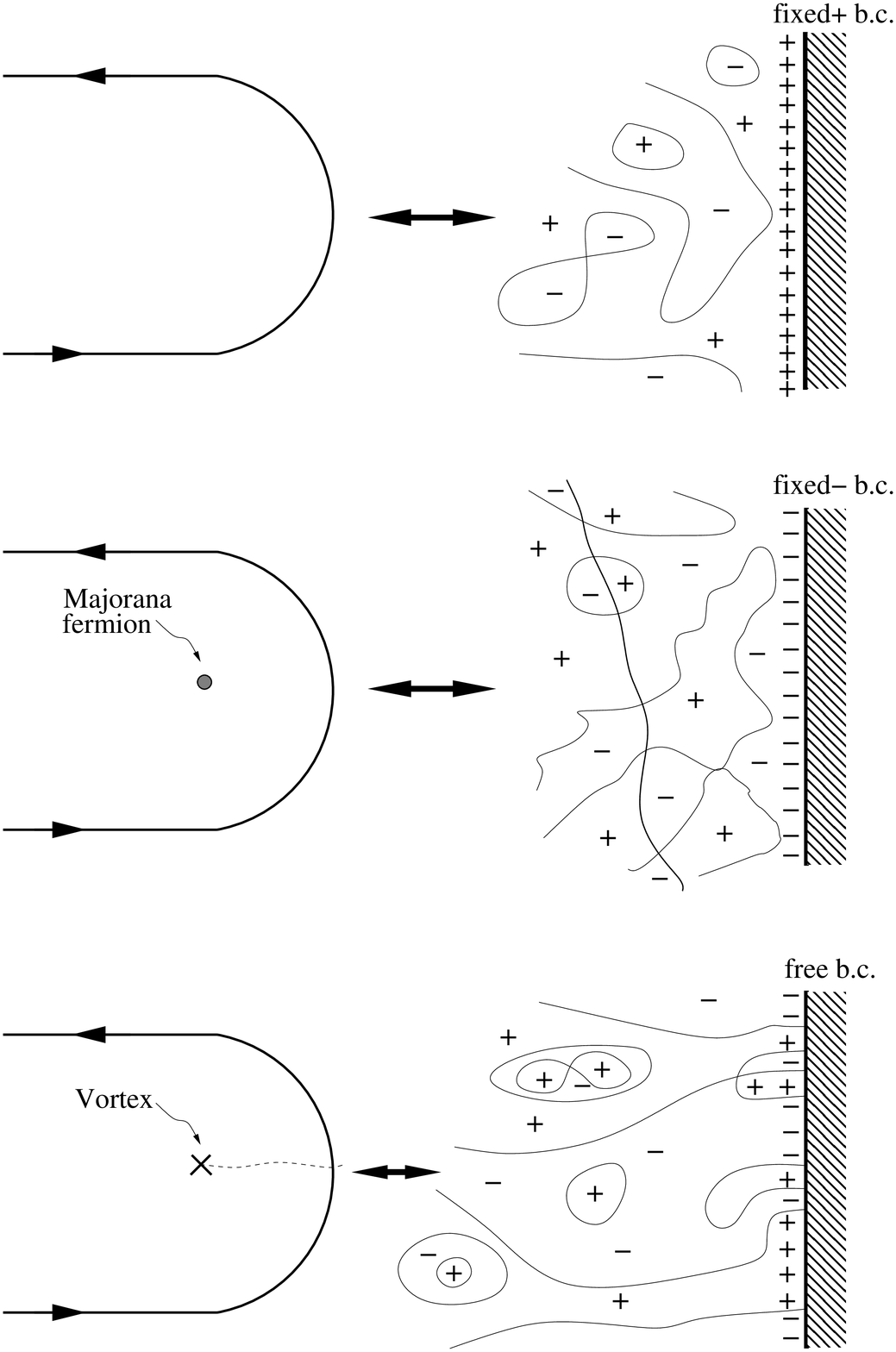}}
\caption{The presence of different bulk excitations in
a $5/2$ quantum Hall droplet is equivalent to different
conformal boundary conditions in the critical Ising
model. In all three cases, the boundary conditions on the other side
of the strip
are taken to be fixed up.}
\label{fig:Majorana-Ising}
\end{figure}

We have thus shown that when the boundary conditions at the end of the
strip are both fixed, this corresponds in fermion language to having
no vortices in the bulk. 
When one boundary condition is free and the
other fixed, this corresponds to having a single vortex.
These situations are illustrated schematically in figure
\ref{fig:Majorana-Ising}.  

Note that there is some ambiguity in the
translation from the fermionic edge theory to the Ising
model. In the latter case, it is clear which end is fixed
and which one is free while, in the former,
a $Z_2$ gauge transformation can exchange the free and fixed
ends (or even put branch cuts in the middle of the strip). 
The reason for this is the following. The gauge transformation
which exchanges the free and fixed ends of the strip
is ${\psi_R}\rightarrow-{\psi_R}$, ${\psi_L}\rightarrow{\psi_L}$.
In Ising language, however, this is simply Kramers-Wannier
duality since it flips the sign of the energy operator
$\epsilon={\psi_R}{\psi_L}$ and, under duality, free
and fixed boundary conditions are switched. A free boundary
condition for the order field is a fixed boundary condition
for the disorder field and vice versa.

Another subtlety is that in the Ising model, there are actually two
types of fixed boundary conditions: all boundary spins up, and all
boundary spins down. These can be understood in the fermion problem by
recalling that the fermion operator creates a cut in the spin field
$\sigma$ (the operator product $\psi(z)\sigma(0)\sim z^{-1/2}
\sigma(0)$).  When boundary conditions are fixed-up on one side of the
strip and fixed down on the other, there must be an odd number of
domain walls stretching throughout the system. When they are fixed the
same on both sides, there are an even number of domain walls. Thus
these two possibilities correspond respectively to odd and even
fermion numbers at the edge. If the total number of electrons
is fixed to be even, then the fermion number at the edge can only
be changed by breaking a pair and putting one fermion at the
edge and the other in the bulk, so even/odd fermion
numbers in the bulk correspond to even/odd fermion
numbers at the edge.

Acting with the vortex creation operator in the bulk changes either of
the two fixed boundary conditions into free.
Since a vortex is a non-Abelian quasiparticle, with quantum dimension
$d=\sqrt{2}$, the free boundary condition has a higher entropy than
fixed by $\ln\sqrt{2}$.  We show in the next section
\ref{sec:free-fixed-flow} that in the case of a vortex in the bulk,
bulk-edge coupling can cause the system to flow from free boundary
condition back to fixed.  This lifts the $\sqrt{2}$-fold degeneracy of
the vortex, leading to an entropy drop $\ln\sqrt{2}$.  On the other
hand, acting with a fermion creation operator leaves free boundary
conditions invariant, and so does not change the entropy. Moreover,
the fixed boundary condition is stable to bulk-edge coupling.  This is
what we expect since a Majorana fermion is an Abelian quasiparticle,
i.e.\ it has quantum dimension $d=1$; therefore fixed up and fixed
down boundary conditions have the same entropy and are stable to
perturbations.

Turning to the case of two vortices in the bulk, the fermions should
again have anti-periodic boundary conditions around the unsquashed
droplet.  It would thus seem that we could either take both ends fixed, or
both ends free, since these are related, in fermionic language, by a
gauge transformation. However, they are not quite physically
equivalent.  If both ends are taken fixed, then the fermion
number parity on the edge is also fixed -- either to $0$ or $1$,
depending on whether the spins on the two ends of the strip are fixed to the
same or different values, respectively. If both ends are taken to be
free, then the fermion number parity is not taken to be fixed and both
$0$ and $1$ are allowed. Since the fermion number parity at the edge
is equal to the value of the qubit formed by the two vortices, we
conclude that both ends fixed is appropriate to the situation in which
the qubit has a fixed value (in this basis) while both ends free is
correct when the qubit does not have a fixed value and is in
an entropy $\ln 2$ mixed state. We will discuss
this further in section \ref{sec:many-qps} and generalize these
results to an arbitrary number of vortices.

\section{Coupling of the edge to a bulk vortex
and the flow from free to fixed Ising boundary conditions}
\label{sec:free-fixed-flow}

In this section and the following ones, we shall focus on various
``tunneling" perturbations which act on the Majorana fields at
$x=0$. The first case we study is the effect of bringing a bulk vortex
close to an edge.  For simplicity, let us suppose that there is only a
single vortex in the bulk.  A vortex has a single Majorana zero mode
localized at its core, which we denote as $\psi_0$.  Since the edge
has a zero mode as well (recall that when there is a single vortex in
the bulk, the boundary conditions at $x=0$ and $x=L$ are the same),
Majorana fermions can tunnel from the vortex to the edge. To study
this within our boundary approach, we choose the point along the edge
at which the tunneling occurs (i.e.\ the closest point to the vortex)
to be one of the boundaries of the squashed system, say that at
$x=0$. The effect of the vortex on the edge then occurs entirely at
$x=0$, so it is convenient to place the branch cut associated with the
vortex there as well.  Thus in the absence of tunneling, we have
$a=b=1$ in (\ref{bcfermion}); in Ising language we have a free
boundary condition at $x=0$ and a fixed boundary condition at $x=L$.

The zero-mode tunneling term in the Lagrangian resulting from
bulk-edge coupling is therefore \cite{Bolech07,Rosenow08}
\begin{equation}
L_h =
i \psi_{0} \partial_t \psi_{0} + 
i h\, \psi_{0} [\psi_R(0) + \psi_L(0)]\ ,
\label{Lresonant}
\end{equation}
where $h$ is the amplitude for tunneling between the edge
and the zero mode associated with the vortex.
Note that the relative sign between the two terms in
the square brackets is consistent with the boundary condition
$a=1$, i.e.\ $\psi_R(0)=\psi_L(0)$, when $h=0$.
The $i$ in front of the coupling to the vortex is necessary
in order for the Hamiltonian to be hermitian, with $h$ real.
The magnitude of $h$ is determined by the distance between
the vortex and the edge; at large distance $r$ from the edge, it should be
$\sim e^{-\Delta r/v}$ where $\Delta$ is the bulk energy gap for
Majorana fermions (which might be smaller than the charge gap
in the $5/2$ quantum Hall state) and $v$ is their velocity.
We will comment below on the sign of $h$.

Since even with the perturbation (\ref{Lresonant}) the action remains
quadratic in the fermions, $\psi(x)$ and
$\psi_0$, it can easily be solved exactly. The equations of motion
for $\psi_0$, $\psi_R$ and $\psi_L$ at
$x=0$ become
\begin{eqnarray*}
2\partial_t \psi_0 &=& h[\psi_R(0) + \psi_L(0)]\ , \\
v_n \psi_R(0) &=& v_n \psi_L(0) + h\psi_0\ , \\
v_n \psi_L(0) &=& v_n \psi_R(0) - h\psi_0 \ .
\end{eqnarray*}
Going to frequency space gives then
\begin{equation}
\psi_R(x=0,\omega) =   \frac{\omega + i\omega_0} {
\omega - i \omega_0} \cdot \psi_L(x=0, \omega) ,
\label{bcres}
\end{equation}
where the scale $\omega_0 \equiv h^2/2v_n$ grows with the bulk-edge
coupling. Intuitively, it is helpful to view this as a scattering
problem. Because of the quadratic Hamiltonian, each incident Majorana
fermion is reflected one-by-one from the boundary, with an 
energy-dependent scattering amplitude given by the phase in
(\ref{bcres}).

\begin{figure}[ht]
\centerline{\includegraphics[width=85mm]{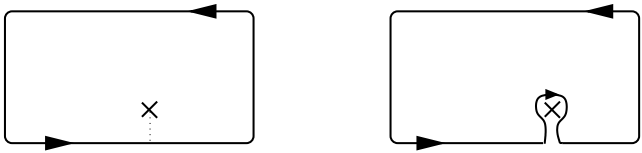}}
\caption{Majorana fermion tunneling between the gapless
chiral edge mode and a bulk quasiparticle/vortex causes the
bulk zero mode to be absorbed by the edge.}
\label{fig:bulk-edge}
\end{figure}

The boundary condition (\ref{bcres}) smoothly interpolates between the
two boundary conditions we discussed previously.  When there is no
edge-vortex coupling ($h=\omega_0=0$), we recover the $a=1$ boundary
condition arising from a single bulk vortex. In the strong-coupling
${\omega_0}\to\infty$ (or, equivalently, DC) limit, we obtain the
boundary condition $a=-1$. This is the boundary condition in the {\em
absence} of the vortex. Thus the presence of the relevant coupling
between vortex and edge causes the cut to ``heal: the zero mode at
the vortex core is effectively absorbed into edge, and annihilates the
edge zero mode.  This situation is sketched schematically in Figure
\ref{fig:bulk-edge}. 
In the language of boundary field theory, the relevant
coupling causes a flow of boundary conditions from $a=1$ to $a=-1$.

In dynamical terms, when a single bulk vortex is brought close to
the droplet edge at time $t=0$, it is scattered strongly by the edge
modes.  At short times $t \ll\omega_0^{-1}$, it has little effect on the
edge modes, but for $ t\gg\omega_0^{-1}$ a $\pi$ phase shift in the
Majorana field is induced at the boundary, changing $a=1$ to $a=-1$.
Another important time scale is set by the system size, $t_L  \equiv L/v_n$,
ie. the time it takes for an edge disturbance to circumnavigate the droplet.
For $t << t_L$ the induced phase shift can be viewed as a ``local" change
and can have no influence on the
other boundary at $x=L$.     However, in the d.c. limit, $t \rightarrow \infty$
(taken before the thermodynamic limit, $t_L \rightarrow \infty$)
the net effect of the crossover
induced by coupling the vortex to the edge is a global change in 
boundary conditions on circumnavigating the droplet from periodic to
anti-periodic. This change in boundary conditions disallows the zero
mode: the vortex is no longer there as far as the edge is concerned.
Of course, in the d.c. limit the point at which the fermion changes sign (in order
to satisfy the anti-periodic boundary condition) can be placed
wherever we like by a (static) gauge transformation; $x=0$ is the most
convenient choice in the squashed geometry.

%large the incoming and outgoing particles, 
%$\psi_R^{out/in} = \psi_R(x=0^\pm,0)$, suffer a simple $\pi$ phase shift: 
%\begin{equation}
%\psi^{out} = - \psi^{in} .
%\end{equation}
%It will be convenient to define a single particle $S-$matrix which relates
%the incident and scattered amplitudes,
%\begin{equation}
%\psi_\alpha^{out} = {\cal S}_{\alpha \beta} \psi_\beta^{in}  ,
%\end{equation}
%with $\alpha,\beta = R,L$.
%In the present situation with a single pinned impurity vortex 
%near the right edge one has,
%\begin{equation}
%{\cal S}_{R,imp} = -  \sigma^z  .
%\end{equation}

A notable property of this crossover is that the phase shift in the DC
limit is independent of the impurity strength $h$.  This should be
contrasted with the analogous problem of a chiral Dirac fermion,
$\chi(x)$, scattering from an impurity level, with Lagrangian
\begin{equation}
{\cal L}_{imp}^D = i d^\dagger \partial_t d + \lambda (\chi^\dagger(0) d + h.c.) + \epsilon_0 d^\dagger d,
\end{equation}
where $d$ and $d^\dagger$ annihilate and create a particle on the
resonant level. 
Here in the DC limit after squashing, $\chi_L(0) 
= e^{i\theta} _{L}\chi_R(0)$ with $\theta = 2
\tan^{-1}(\omega_0/\epsilon_0)$ and $\omega_0=\lambda^2/2v_n$. 
It is only when the impurity energy is fine-tuned to zero, i.e.\ on resonance,
that the phase shift for complex fermions becomes independent
of the scattering strength, $\theta(\epsilon_0=0) = \pi$, as in the
Majorana case.
Since $\psi_0^2 = 1$, a chiral Majorana fermion is {\it always}
on-resonance, and no fine-tuning is required.    
A single Majorana fermion
is not a physically meaningful object - they can only exist in pairs.

The absorption of the Majorana zero mode by the edge leads to a loss
of entropy.  For the case of Dirac fermions tuned to resonance,
$\epsilon_0=0$, the entropy loss is simply $\ln 2$ since the
$d-$level constitutes a two-level system ($d^\dagger d=0,1$) which is
screened by the edge fermions.  The entropy loss for the Majorana
fermion case can be inferred by introducing a second identical copy of
the Majorana edge plus impurity system.  It is convenient to introduce
a complex Dirac fermion for both the edge and impurity as $\chi =
(\psi^\alpha + i \psi_\beta)/\sqrt{2}$ and $d = (\psi_0^\alpha + i
\psi_0^\beta)/\sqrt{2}$, where $\alpha,\beta$ refer to the two copies.  When
re-expressed in terms of these complex fermions, the total Lagrangian
${\cal L}_\alpha + {\cal L}_\beta$ becomes identical to the Dirac case with
$\epsilon_0$.  Since the two Majorana copies are decoupled, the
entropy drop is additive.  Thus, the entropy change during the
crossover induced by coupling the single Majorana zero mode to the
edge is $\Delta S = - \frac{1}{2}\ln 2 = - \ln \sqrt{2}$.

Since the model is quadratic in fermions, much more than the entropy
can be computed. In fact, the full partition function and explicit
correlation functions have been computed
along this whole flow by using the ``boundary state''
\cite{Chatterjee94}. 

The vortex-edge coupling $h$ corresponds to a boundary magnetic
field in the Ising language. This initially may seem strange, since a bulk
magnetic field couples to the spin field $\sigma$, not the fermion
$\psi$. However, the spin field in the Ising model is a product of
left and right components of $\sigma$. At a boundary, one must
identify these components of $\sigma$ just like we identified $\psi_L$
and $\psi_R$ in (\ref{bcfermion}). Thus at the boundary, the product
$\sigma_L\sigma_R$ becomes the fusion $\sigma(0)\times \sigma(0) = I +
\psi(0)$, so that a boundary magnetic field indeed couples to $\psi$
as in (\ref{Lresonant}) \cite{Cardy89}.
This picture in terms of the Ising model gives useful intuition.
When the boundary magnetic field gets large ($|h|\to\infty$),
the boundary spin is fixed to $+1$ or $-1$, depending on the sign of
$h$. Thus in Ising language, the coupling causes a flow from
free to fixed boundary conditions; it has long been known that this
causes a change $\Delta S=-\ln\sqrt{2}$ in entropy
\cite{Cardy89,Affleck91}. It is also clear what sign $h$
should have. If the total fermion number is even, then when the
boundary at $x=0$ flows to fixed boundary condition, the Ising
spin must be fixed to the same value as at $x=L$; if the fermion
number is odd, it must be fixed to the opposite value. Since
the sign of $h$ determines the sign of the Ising spin at $x=0$,
its sign is determined by the fermion number parity
and the boundary condition at $x=L$.

For a boundary magnetic field to have any effect,
we must start, at $h=0$, with free boundary conditions, i.e.\ $a=1$.
If the boundary condition is fixed, then coupling to a
boundary magnetic field can have no effect:
there is nothing to couple to, since the boundary spin is not allowed
to flip. This can also be seen from (\ref{Lresonant}): if
${\psi_R}(0)=-{\psi_L}(0)$, then this term vanishes.
In order to have a free boundary condition at $x=0$,
there must be present the Ising analog of a vortex, which in this context is called
the ``twist'' operator \cite{Ginsparg}. (The twist operator turns out,
not surprisingly, to be the spin field.) The resulting zero mode is
$\psi_0$. $\psi_0$ can also be viewed from a formal perspective
as the Klein field necessary to make the second term in
the Lagrangian (\ref{Lresonant}) bosonic.
Finally, it is also worth mentioning that the Majorana crossover is formally
identical to that of an anisotropic 2-channel Kondo problem at its
Toulouse point; the Majorana zero mode operator $\psi_0$
is mapped to the $\sigma^x$ operator for the Kondo spin \cite{Emery92}.

\section{Point contacts and boundary conformal field theory}
\label{sec:point-contact}

In this section we explain how to understand the possible critical
behavior of topological states in the presence of a point contact.
The point contact allows backscattering of right movers on the top to
left movers on the bottom. We use `backscattering' to denote tunneling
from one edge to the other across the point contact; we will use
`tunneling' generically to describe tunneling from the edge to a bulk
quasiparticle or to another edge and, especially, from one droplet to
another.  Generically, backscattering destroys criticality, although
in the next section \ref{sec:neutral-pt-contact} we will discuss a
special case where criticality survives even in the presence of
backscattering. Before studying backscattering, however, it is useful to
understand how to deal with a point contact using boundary conformal
field theory.

\begin{figure}
\centerline{\includegraphics[width=45mm]{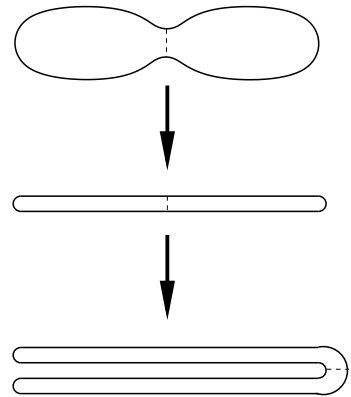}}
\caption{Deforming a topological state with a
point contact into a horseshoe
shape so that it can be mapped onto a boundary problem.}
\label{fig:folded-drop}
\end{figure}

To turn a point contact into a boundary problem, we must
squash and then ``fold'' the system around the point contact, so that
effectively we have two copies of the system coupled at one of their
boundaries \cite{Wong94}.  Namely, we place the point contact somewhere
in the middle of the sample, far from the ends.
Once we have squashed, the point contact is effectively an
impurity in this non-chiral system, or a defect line in the equivalent
two-dimensional classical system. By a conformal transformation, we
can put the point contact/defect line at $x=0$, and take
$x\in[-L,L]$. This impurity/defect problem can in turn be turned into a
boundary problem by folding the system at the impurity.  What
folding means is that the droplet has now been deformed into
a horseshoe shape, with the point contact at the bend in the
horseshoe, as depicted in Fig.\ \ref{fig:folded-drop}. Since edge
interactions are local, the top and bottom parts of the horseshoe are
two copies of the system of length $L$, coupled only via the point
contact.

For the Majorana fermion edge mode, folding
turns out to simplify the problem considerably. The reason is that
even though a single Majorana fermion cannot be bosonized, a pair
can. In fact, this is the easiest way to compute explicit correlators
in the critical Ising field theory: square the correlator, bosonize,
compute the correlator, and then take the square root
\cite{diFrancesco87}.  This procedure is even more natural in our
context, because the folding automatically doubles the degrees of
freedom. Bosonizing the neutral sector of a $\nu=5/2$ point contact
allows to make contact with the detailed results of Ref.\
\onlinecite{Oshikawa97}.

We have seen in section \ref{sec:Majorana-Ising} that for the
Majorana fermion edge mode with no vortices in the bulk, the boundary
conditions in the squashed system are fixed at both ends (in Ising
language), so that $\psi_{L}(-L)=-\psi_R(-L)$ and
$\psi_L(L)=\psi_R(L)$.  Labeling the two halves of the droplet by $1$
and $2$, folding results in two right-moving modes,
$\psi_{1R}(x)={\psi_R}(x)$, $\psi_{2R}(x)={\psi_L}(-x)$ and two
left-moving modes, $\psi_{1L}(x)={\psi_L}(x)$,
$\psi_{2L}(x)=-{\psi_R}(-x)$ with $x\in [0,L]$. The reason for the
extra minus sign in the last of these is that, as explained in section
\ref{sec:Majorana-Ising}, a given boundary condition (free or fixed)
at the right end of the line segment has the opposite sign from the
same boundary condition at the left end (i.e.\ $a=-b$ in
eqn.\ \ref{bcfermion}). Folding exchanges left and right ends, so the
extra sign interchanges boundary conditions appropriately.
Equivalently, we are choosing a gauge in which there is no branch cut
at either end of the strip (which, after folding, translates to fixed
Ising boundary conditions at both ends). In the gauge which we have
chosen, the branch cut which is necessary in the absence of bulk
vortices is at the point contact.  
%Later in this section, we will
%consider the case in which there is a bulk vortex, when there wont be
%a branch cut at the point contact either.

\subsection{Conformal boundary conditions}
\label{subsec:conformal-bc}

A key component of our analysis is to understand the {\em boundary
fixed points}, or in more formal language, the {\em conformal boundary
conditions}. (We use the two descriptions interchangeably.) A fixed
point of the renormalization group is scale invariant, and typically
in two spacetime dimensions, conformally invariant as well.  In a
topological state, the edge is scale and conformally invariant
in the absence of any tunneling.
Once we allow tunneling at a point, scale
invariance is typically broken at that point, but of course still
remains valid in the rest of the edge theory. Thus the situation in the
presence of tunneling is generally a conformal field theory with
boundary conditions breaking the scale and conformal
invariance. However, as we already saw in our analysis of vortex/edge
tunneling in section \ref{sec:free-fixed-flow}, at low temperatures and
frequencies, the boundary conditions effectively flow to a new
boundary fixed point. Thus before exploring which types of tunneling
cause which flows, it is very useful to first understand the
different possible boundary conditions which preserve scale and
conformal invariance of the full system, including the boundary.

Some of the conformal boundary conditions for a point contact in a
Majorana fermion edge mode are fairly obvious from both the Ising and fermionic
points of view: these are ``product'' boundary conditions, in which the
two copies decouple. Physically, this corresponds to the point contact
effectively splitting the system in two. We have already shown that
the boundary condition (\ref{bcfermion}) has $a=-1$ if there are no
vortices in the bulk, and $a=1$ if there is a single vortex. Thus, in
the two copies, there are four possibilities: $(a_1,a_2)=(1,1)$,
$(-1,1)$, $(1,-1)$, and $(-1,-1)$. In the Ising language, $a=1$ and
$a=-1$ correspond to ``free and ``fixed'' respectively. Thus if the
point contact splits the system in two, the boundary conditions
at $x=0$ in the folded system are
(fixed, fixed), (free, fixed), (fixed, free), or (free, free),
depending on if and where vortices are located. (As discussed in section
\ref{sec:Majorana-Ising}, there are two possibilities for each fixed
boundary condition.)

Obviously, product boundary conditions are not the end of the
story. For example, one boundary condition corresponds to no defect at
all, i.e.\ no backscattering occurring at the point contact. In this case,
the point contact is effectively not there at all. Of course, we are
still free to fold the system at this point, and treat the model as a
boundary problem.  We refer to the resulting boundary condition as
``transmitting''. Before folding, this boundary condition simply
ties the left-moving fields in the two copies together at $x=0$. After
folding, left movers in copy 2 become right movers, so 
\begin{equation}
\hbox{transmitting: } 
\psi_{1L}(0)=\psi_{2R}(0),\quad \psi_{1R}(0)=-\psi_{2L}(0)\ .
\label{bctransmit}
\end{equation}
The minus sign in the latter term arises from the extra minus
sign in the folding discussed above.  This boundary condition
(\ref{bctransmit}) results from the equations of motion of the
action (\ref{eqn:strip-action}) plus the
boundary term
\begin{equation}
L_{\hbox{transmit}}= i{v_n}\left(\psi_{2L}(0)\psi_{1R}(0)
-\psi_{1L}(0)\psi_{2R}(0)\right)\ .
\label{Ltransmit}
\end{equation}
This term (\ref{Ltransmit}) is not the only quadratic boundary term
one can add. In section \ref{sec:neutral-pt-contact}, we discuss how
allowing fermion tunneling across the point contact results in a line
of boundary conditions.

Understanding transmitting boundary conditions allows us to apply all
our results to a topological state on an {annulus} as well as a
disc. We have seen that after squashing and folding, the disc becomes
a doubled system with the two copies coupled at $x=0$, the location of
the point contact. The boundary conditions at the far end $x=L$ turn
into (fixed, fixed) boundary conditions, leaving the two copies
decoupled there.  For an annulus, we can still keep the point contact
at $x=0$, but now put {transmitting} boundary conditions at
$x=L$. This sews the top edges of the two copies together, and the
bottom edges together, but does not sew top to bottom. In the original
unfolded case, this indeed corresponds to an annular geometry. Obvious
physical arguments indicate that this change of boundary conditions at
$x=L$ should have no effect on the behavior of the point contact, and
indeed detailed computations confirm this. However, this change can
and does effect global properties like the total entropy, and we will
return to the annulus at the end of section \ref{sec:flows}.

\subsection{Bosonization}

To understand the conformal boundary conditions for the point contact
in the Majorana fermion problem in more depth, it is useful to
bosonize the fields. In fact, using bosonization, all the boundary
fixed points for two Ising conformal field theories coupled at the
boundary have already been found \cite{Oshikawa97}. This will enable
us to show that all these conformal boundary conditions (and then
some) can be obtained in our problem, once we allow vortices to be
present.  We thus conclude this section by outlining some results of
Ref.\ \onlinecite{Oshikawa97}.

After folding, we have two Majorana fermions $\psi_1$ and $\psi_2$,
coupled only at the boundary $x=0$.  A single Dirac fermion $\chi$ can
be formed out of these two Majorana fermions in the same way a complex
number is formed out of two reals.  A free Dirac fermion $\chi$ can be
bosonized according to $\chi_{R}=e^{i\varphi_{R}}$ and
$\chi_{L}=e^{i\varphi_{L}}$, where we normalize the bosonic fields so
that the scaling dimension of $e^{i\alpha\varphi_L}$ or
$e^{i\alpha\varphi_R}$ is $\alpha^2/2$.  It is often useful to combine
the chiral bosons into a single boson
$\varphi=\frac{1}{2}(\varphi_L+\varphi_R)$. In this Ising model,
$\varphi$ has radius $1$, meaning that we identify
$\varphi\sim\varphi +2\pi$. This is tantamount to saying that
restricting $\alpha$ to be an integer results in only fermionic or
bosonic operators.  The Lagrangian for the folded system in the
bosonic picture is then
\begin{equation}
\label{eqn:bosonized}
L_0=\frac{1}{2\pi}\int_0^L dx
\left[\left(\frac{\partial\varphi}{\partial t}\right)^2 - {v_n^2}
\left(\frac{\partial\varphi}{\partial x}\right)^2 \right]\ .
\end{equation}
The Lagrangian can equivalently be written in terms of a dual boson
$\widetilde{\varphi}=\frac{1}{2}({\varphi_R}-{\varphi_L})$.
In the Ising model, 
$e^{i\varphi/2}$ is the product of the two Ising spin
fields, while $e^{i\widetilde{\varphi}/2}$ is the product of the two
disorder fields.

Classifying the various possible conformally-invariant
boundary conditions of the bosonic model (\ref{eqn:bosonized}),
which include the different product boundary conditions
as well as (\ref{eqn:contin-Diri-bc}), is equivalent to classifying
the different possible fixed points of the critical Ising
model with a defect line. Indeed, Oshikawa and Affleck \cite{Oshikawa97}
analyzed this model by folding the Ising model about the defect,
precisely as we folded our droplet above, and then bosonizing
the two resulting copies of the Ising model.
All of the conformal boundary conditions
were found. Moreover, the boundary states were constructed
explicitly, which allows correlators to be computed exactly for any
conformal boundary condition. 

The possible conformal boundary conditions are summarized in the table
\ref{tab:OA}.  There are two different lines of boundary fixed points,
dubbed `continuous Dirichlet' and `continuous Neumann' in Ref.
\onlinecite{Oshikawa97}. The former corresponds to setting the field
$\varphi(0)=\varphi_0$ at the boundary, while the latter corresponds
to setting the dual field
$\widetilde{\varphi}(0)=\widetilde{\varphi}_0$.  The remaining
boundary conditions are of product type: either (fixed, fixed), (free,
fixed), or (fixed, free). 
Table \ref{tab:OA} also lists the contribution of each type of
conformal boundary condition to the entropy; we discuss these values
in depth in the section \ref{sec:flows}.

\begin{table}
\begin{center}
\begin{tabular}{c|c|c}
 {\hskip 0.4 cm} Boundary condition {\hskip 0.4 cm} & {\hskip 0.4 cm} parameter
  {\hskip 0.4 cm}  & {\hskip 0.4 cm} entropy {\hskip 0.4 cm}\\ \hline
continuous Neumann & $2\widetilde{\varphi}_0={{\delta}} + \frac{\pi}{2}$ & $\ln\sqrt{2}$ \\
continuous Dirichlet & $2{\varphi_0}={\delta} + \frac{\pi}{2}$ & 0 \\
(free, fixed)& & $-\ln\sqrt{2}$\\
(fixed, fixed)& & $-\ln 2$\\
\end{tabular}
\end{center}
\caption{Summary of boundary conditions}
\label{tab:OA}
\end{table}

The continuous Dirichlet (CD) line of fixed points is easy to
understand in the language of the classical Ising model with a
defect. To move the model off of criticality, one varies the coupling
between adjacent spins. In field theory, this corresponds to adding
the energy operator $\epsilon(x,t)$ to the Lagrangian density. This
operator turns out to have dimension one, so it is indeed a relevant
perturbation in the bulk. However, if we add
$\epsilon(x=0,t)$ to the original (unfolded) quantum Ising chain
at a single point $x=0$, this is an exactly marginal perturbation. In
the two-dimensional classical field theory, this corresponds to a defect line
at $x=0$ for all $\tau$. In the 2d classical lattice model, this amounts to
a defect with deformed couplings between the adjacent spins across the
defect. Varying $\varphi_0$ to move along the CD line
therefore corresponds to varying the Ising coupling at a single link in the
quantum transverse field model, or along the defect line in the
classical lattice model.

The product boundary conditions (free, free) do not
appear separately in the table, because they are a particular point on
the CD line.  This is obvious from the Ising defect
interpretation: taking the limit of zero link coupling along the
defect splits the system in two, and puts free boundary conditions on
the spins on each side of the defect.

We will describe in detail in the
next section how  continuous
Dirichlet boundary conditions correspond to allowing Majorana
fermion backscattering at the point contact.
We will also show  there that 
the continuous Neumann (CN) boundary conditions
arise in a topological state when a vortex is pinned in the bulk.

%, with the identification
%${\varphi_0}=\frac{\delta}{2}+\frac{\pi}{4}$.

\section{Majorana fermion backscattering at a point contact}
\label{sec:neutral-pt-contact}

In this
section, we discuss in depth the simplest kind of backscattering at a
point contact,
that of Majorana fermions.
This backscattering is identical for  $p+ip$ superconductor,
its quantum-disordered counterpart, and
$\nu=5/2$ fractional quantum Hall effect, 
because in the latter the Majorana fermion has no charge,
so that the extra bosonic field does not affect its tunneling.

%defer the general case to section
%\ref{sec:general-case}. 

The effects of backscattering a point contact can be treated as
different boundary conditions at $x=0$ in the folded picture. For
Majorana fermion backscattering, these boundary conditions remain
conformal.  The reason is that free fermionic fields have dimension
$1/2$, so any bilinear $\psi_1\psi_2$ has dimension 1.  Such a
bilinear backscattering operator at the point contact is an exactly
marginal boundary operator, unlike the relevant backscattering in a
Luttinger liquid \cite{Kane92,Fendley95}.  Tuning the strength of the
backscattering results in a {\em line} of conformal boundary
conditions. Such a line does not occur in the case of a single
Majorana fermion (e.g.\ at one of the ends of the strip): because of
the boundary condition (\ref{bcfermion}), a bilinear
$\psi_L(0)\psi_R(0)$ can be fused together, yielding the identity
operator and thus a trivial boundary perturbation. Since the point
contact results (after folding) in two copies of a Majorana fermion
with a boundary, the bilinears $\psi_{1L}(0)\psi_{1R}(0)$ and
$\psi_{2L}(0)\psi_{2R}(0)$ are not trivial.

As we saw from the bosonization analysis, there are in fact two
critical lines. We show in this section precisely how to move along
either of these critical lines by tuning the backscattering
strength across the point contact.
The continuous Dirichlet line (of which
transmitting and (free,free) are special cases) arises when there are
no vortices present. The continuous Neumann line occurs when there is
a single vortex pinned in the bulk.

\subsection{Fermion backscattering in the absence of vortices}

We start our analysis of fermion tunneling by considering a disk with
no vortices. With no tunneling of any sort, the folded model
has transmitting boundary conditions at $x=0$. Allowing backscattering
there perturbs the
transmitting boundary condition by adding the operator
\begin{equation}
{L}_f = i\lambda_f(\psi_{1L}(0)\psi_{1R}(0) +
\psi_{2L}(0)\psi_{2R}(0))
\label{Lm}
\end{equation}
to the Lagrangian. Because the
transmitting boundary condition (\ref{bctransmit}) relates the two
terms in (\ref{Lm}), fermionic anticommutation requires a relative
plus sign to obtain a non-trivial perturbation.

\begin{figure}[h]
\centerline{\includegraphics[width=85mm]{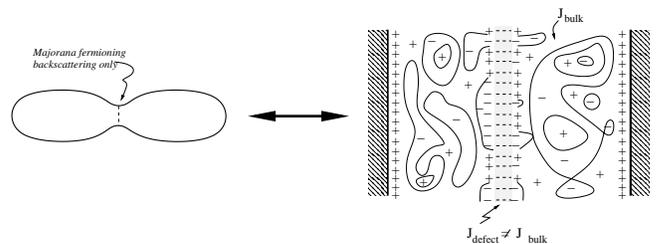}}
\caption{A point contact at which Majorana fermion backscattering
is the only non-zero tunneling process is equivalent to
the Ising model with a column of bonds at which
$J_{\rm defect}\neq J_{\rm bulk}$.}
\label{fig:contact-Idefect}
\end{figure}

Since $L_f$ is exactly marginal, adding it to the Lagrangian results
in a line of boundary fixed points parametrized by $\lambda_f$. This
line is precisely the continuous Dirichlet line discussed at the end
of the previous section. This is obvious from the Ising defect
interpretation. Transmitting boundary conditions correspond to no
defect at all, so in the lattice model this amounts to setting the
link coupling across the defect to be the same as the link coupling
everywhere else. This, therefore, is a point on the CD line. As is well
known, the marginal energy operator $\epsilon$ is $\psi_L\psi_R$ in
fermionic language. Perturbing by $L_f$ is therefore the same as varying
the link coupling across the defect, as illustrated schematically
in figure \ref{fig:contact-Idefect}. Thus, varying the Majorana fermion
backscattering $\lambda_f$ at the point contact corresponds to to
varying $\varphi_0$ to move along the CD line.  Below, we relate
$\lambda_f$ to ${\varphi_0}$ explicitly.

Since the boundary perturbation is quadratic, we can derive the exact
equations of motion.  The action of the folded system is of the form
(\ref{eqn:action-with-b}), where now ${\cal L}_{\rm edge}$ is
comprised of two copies of (\ref{eqn:fermion-Lagrangian}) for the two
non-chiral Majorana fermions, and the boundary terms are
(\ref{bctransmit}) and (\ref{Lm}). The solution can be expressed most
simply in terms of the Dirac fermions ${\Psi_R}=\psi_{2R}+i\psi_{1R}$
and ${\Psi_L}=\psi_{1L}-i\psi_{2L}$. The effect of the point contact
is simply a phase shift:
\begin{equation}
\label{eqn:phase-shift}
{\Psi_R}(0) = e^{i\delta}\,{\Psi_L}(0)
\end{equation}
where $\tan(\delta/2) = {\lambda_f}/2{v_n}$.  (Another way to solve
the system is to unfold and redraw the point contact with the
left-movers flipped so that they became right-moving. Then the two
right-moving Majorana fermions can be combined into a Dirac fermion,
and again the effect of (\ref{Lm}) is simply a
phase shift.)

It is useful to re-express this in terms of a {\em reflection matrix}
${\cal R}$, which describes how the Majorana fermions behave when they
bounce off the boundary.  In the original, pre-folding, picture, this
is the scattering matrix off the point contact. Since the Majorana
fermions are real, the reflection matrix is necessarily real.  One
situation we have already discussed is transmitting, i.e.\ when there
is effectively no point contact at all. In the folded system, perfect
transmission (\ref{bctransmit}) corresponds to reflection matrix
\begin{equation}
{\cal R}^{\rm transmit} = -i\sigma^y\ .
\label{Rtransmit}
\end{equation}
Since the action remains quadratic in
terms of the Majorana fermions, one can compute the reflection matrix
directly or rewrite the solution from the Dirac fermion, yielding
\begin{eqnarray}
\label{R-general}
{\cal R}(\lambda_f) &=&
\cos\delta \, (-i\sigma^y) + \sin\delta \, \mathds{1}
\cr
 &=& \frac{1-({\lambda_f}/2{v_n})^2}{1+({\lambda_f}/2{v_n})^2}
\, (-i\sigma^y) 
+ \frac{{\lambda_f}/{v_n}}{1+({\lambda_f}/2{v_n})^2}\, \mathds{1}\cr
& &
\end{eqnarray}
where $\lambda_f$ controls the amount of backscattering.
When $\lambda_f$
is zero, we recover the transmitting case (\ref{Rtransmit}).

To make contact with the conventions of Oshikawa and Affleck
\cite{Oshikawa97}, the Majorana fermions need to be combined in a
slightly different fashion than the $\Psi$ defined above. To be
precise, we write $e^{i\varphi_R}=\chi_{R}=-\psi_{1R}+i\psi_{2R}$ and
$e^{i\varphi_L}=\chi_{L}=\psi_{1L}+i\psi_{2L}$. Then, in terms of the
Dirac fermion $\chi$, the boundary conditions (\ref{eqn:phase-shift})
or, equivalently, (\ref{R-general}) take the form ${\chi_R}(0) = i
e^{i\delta} {\chi_L^\dagger}(0)$.  The Dirichlet boundary condition
$\varphi(0)=\varphi_0$ discussed at the end of the last section then
relates $\delta$ and $\varphi_0$ as \cite{Quella07}
\begin{eqnarray}
\label{eqn:contin-Diri-bc}
\varphi(0)&=&\frac{\delta}{2}+\frac{\pi}{4}\cr
&=&\tan^{-1}\!\left(\frac{2{v_n}+{\lambda_f}}{2{v_n}-{\lambda_f}}\right)\ .
\end{eqnarray}

%We already took advantage of the equivalence between two
%Majorana fermions and a Dirac fermion in the previous
%subsection when we expressed the effect of non-resonant
%Majorana fermion backscattering in terms of a phase shift $\delta$.

% can be traced to
%the peculiar minus sign in the first term on the right-hand-side
%of the definition ${\chi_R}=-\psi_{1R}+i\psi_{2L}$.
%This minus sign is simply a $Z_2$ gauge choice which
%we have made in order to
%make contact with the results of Ref. \onlinecite{Oshikawa97}
%on the Ising model (where the boundary conditions at $x=L$
%are different because they considered the Ising model
%with periodic boundary condition, rather than fixed at both ends of
%the strip). We certainly could have made a different gauge choice,
%and defined $\chi$ differently, in which case the U(1) symmetry
%of the Dirac fermions would have been preserved.

%As we increase $|\lambda_f|$, we increase
%the backscattering across the point contact. 

The backscattering strength $\lambda_f$ therefore parametrizes
a {\em line} of critical boundary conditions. 
To understand this line of critical points, it is useful to
explore some special values of the phase shift $\delta$. 

At ${\lambda_f}=2{v_n}$, we have $\delta=\pi/2$ and $\varphi_0=\pi/2$.
the reflection matrix simply becomes the identity matrix $\mathds{1}$;
the two copies are no longer coupled.  
From (\ref{R-general}), we see that each resulting
droplet has boundary condition (\ref{bcfermion}) with $a=1$, i.e.
\begin{equation}
\hbox{(free, free):  }\  
\psi_{1R}(0)=\psi_{1L}(0),\:\: \psi_{2R}(0)=\psi_{2L}(0)\ .
\label{free-free}
\end{equation}
In Ising language, this is the product boundary condition (free,
free).  
%In terms of the Dirac fermion $\Psi$, this is
%$\chi_R(0)=-{\chi_L^\dagger}(0)$, so every fermion incident upon the
%point contact is backscattered across it. 
The system is effectively
split into two. This split by fermion backscattering is smooth,
without a crossover scale. This is unlike the case of vortex
backscattering, which we discussed at length in two recent papers
\cite{Fendley06,Fendley07a} and will review in section \ref{sec:flows}
below.  Also unlike vortex backscattering, fermion backscattering
results in droplets with free boundary conditions at $x=0$ and fixed
boundary conditions at $x=L$. Thus a pair of vortices has effectively been
nucleated at the point contact, so that each of the two droplets then
has a zero mode.

%In bosonic language \cite{Oshikawa97}, this translates to $\varphi_0=\pi/2$. 

At ${\lambda_f}=-2{v_n}$, $\delta=-\pi/2$ and ${\varphi_0}=0$,
so the reflection matrix is
$-\mathds{1}$. %We now have $\chi_R(0)={\chi_L^\dagger}(0)$, Again,
every fermion is backscattered, but this time neither droplet has a
zero mode. One might be tempted to conclude that this is (fixed,
fixed) boundary conditions, but this is not quite right. Changing the
strength of Ising coupling or, equivalently, adding a fermion bilinear
to the action cannot favor either up-spins or down. Hence, the
boundary conditions of the two droplets cannot be fixed. When
$\varphi_0=0$, the Ising coupling at the defect line is
infinite. Therefore, the two Ising spins on either side of the defect
are fixed to have the same value; however, this value is equally
likely to be up or down.  In fact, since the Ising coupling at the
defect is infinite while the transverse field remains finite, the
defect spins are not only equally likely to be up or down, but they
are also not flipped by local dynamics.  (In contrast, in the case of
a free boundary condition, the boundary spin still fluctuates as a
result of the transverse field, and only has entropy $\ln\sqrt{2}$.)
Thus, we will call these boundary conditions $(\pm,\pm)$.  They are
almost (fixed, fixed), except that the value to which the two spins
are fixed is a spin-$1/2$ degree of freedom (i.e.\ a two-level
system).

%In subsection
%\ref{subsec:Fixed-fixed}, we will discuss the interpretation of this
%boundary condition in more detail.

Another boundary condition on the Dirichlet line is
${\varphi_0}=3\pi/4$, i.e.\ $\delta=\pi$. There is no backscattering
at the point contact, so it is almost the same as transmitting
boundary conditions, except for one thing: every fermion transmitted
picks up a phase shift of $\pi$. In Ising language, this is an
antiferromagnetic defect at which the Ising coupling is equal in
magnitude to that in the bulk but opposite in sign.  This follows from
the result of Ref. \onlinecite{Oshikawa97}
that flipping the sign of the link
coupling sends $\varphi_0\to \pi-\varphi_0$, so if the magnitude is
unchanged from the transmitting boundary condition
($\varphi_0=\pi/4$), then $\varphi_0=3\pi/4$.

The bosonic formulation of this critical line makes it is possible to
compute exact correlation functions along the edge for any value of
$\delta$. Fermionic correlators are of course trivial to find, but to
find ones involving spin fields requires computing the boundary
state. Detailed expressions can be found in
[\onlinecite{Oshikawa97}]. One interesting result is the dimension
$\Delta_b$ of the spin field along the defect, defined so that the
correlator of two $\sigma$ operators at the point contact at different
times falls off as
$$\left\langle\sigma(0,t_1) \sigma(0,t_2)\right\rangle
 \sim \frac{1}{(t_1-t_2)^{2\Delta_\sigma}}$$
The exact expression is \cite{Oshikawa97}
\begin{equation}
\Delta_b(\delta) = \frac{1}{8}\left(1+\frac{2\delta}{\pi}\right)^2\ .
\label{deltab}
\end{equation}
In Ising language, a boundary magnetic field couples to the spin
field, so $\Delta_b$ is the dimension of the operator coupling to this
field. At the (free, free) point, $\delta=\pi/2$, we find
$\Delta=1/2$. This is not surprising since, as we described in the previous section, in a single Ising model with free boundary conditions,
the left and right components of the spin field fuse to form the dimension-1/2 fermion so that a boundary magnetic field
has scaling dimension $1/2$. At the
transmitting point $\delta=0$, the left and right components of the
spin field are decoupled, so $\Delta_b$ is the sum $1/16+1/16=1/8$.
At the $(\pm,\pm)$ fixed point $\delta=-\pi/2$, so we have
$\Delta_b=0$ here. This is a reflection of the fact that even an
infinitesimal boundary magnetic field will favor either $(+,+)$ or $(-,-)$,
thus splitting these two degenerate levels by
adding a dimension-0 $S_z$ term.

\subsection{A Point Contact with a Localized Vortex
in the Bulk: Continuous Neumann Boundary Conditions}
\label{subsec:contin-Neumann}

The situations discussed above do not exhaust the possibilities for
critical behavior at the point contact. Indeed, as seen in table
\ref{tab:OA}, there is another critical line of conformal boundary
conditions, dubbed ``continuous Neumann'' (CN) in
Ref. \onlinecite{Oshikawa97}.  CN boundary conditions do not occur in
the classical Ising model with a defect line; in the quantum
transverse field Ising model, they correspond to varying the Ising
coupling $J$ at a single link in the presence of a peculiar
${\sigma_{-1}^z}{\sigma_0^x}$ term across this link ($i=-1,0$ are the
sites on either side of the defect link) \cite{Oshikawa97}.  Since the
Ising spin field ${\sigma^z}$ is the vortex creation operator, we
expect that the CN line can be realized in the $p+ip$
superconductor/$\nu=5/2$ quantum Hall state with a single vortex in the
bulk.

To prove our assertion, note that the state
with a vortex in the bulk and no fermion backscattering
has a higher entropy than one without a vortex
by $\ln\sqrt{2}$ (we saw this in section \ref{sec:free-fixed-flow}
by deforming the droplet so that that the vortex is at an
endpoint of the strip and its presence or absence is simply the
difference between free and fixed boundary conditions).
But this is precisely the entropy difference
between the CN
and CD lines \cite{Oshikawa97}.
Since the state without a vortex and no backscattering
is the transmitting point (i.e.\ $\delta=0$ or, equivalently,
${\varphi_0}=\pi/4$) point on the CD line,
we conclude that the state with a vortex and no backscattering --
which has a higher entropy by $\ln\sqrt{2}$ --
is on the CN line. We dub this the
dub the `Neumann-transmitting' point.
We will show in this section that we can move away from this point along
CN line by allowing (non-resonant) fermion backscattering.
%In this subsection, we discuss some special points along the
%CN line and their equivalents in the fermionic picture.
%We then discuss the flow from the CN line
%to the CD line.

In fermionic language, the CN case
is not really substantively different from the
CD case unless the vortex
is pinned right at the point contact and we consider its
coupling to the edge (which leads to a flow from Neumann
to Dirichlet). Otherwise, the effect of the
vortex can simply be absorbed into the boundary conditions
at $x=L$; the boundary condition at the point contact will then be
precisely the same as in the absence of the vortex. However, the translation to
the Ising model and, especially, to the results of Ref. \onlinecite{Oshikawa97}
can be made more directly if we keep the Ising boundary conditions
in both copies fixed at $x=L$. Furthermore, this formulation of the
problem allows us to consider the interesting situation in which
the vortex is in the point contact and Majorana fermions can (resonantly)
tunnel between the vortex and {\em both edges}. We discuss this in the
next section; for the rest of this section we discuss fermion
backscattering and the CN line.

In the unfolded picture, we have periodic boundary conditions,
so that $a=b=1$, i.e.\ free at $x=-L$ end and fixed at $x=L$.
We now fold the strip in half, as before, but now we define
the two right-moving modes as $\psi_{1R}(x)={\psi_R}(x)$,
$\psi_{2R}(x)={\psi_L}(-x)$, and the two left-moving modes as
$\psi_{1L}(x)={\psi_L}(x)$, $\psi_{2L}(x)={\psi_R}(-x)$ with
$x\in [0,L]$. The absence of a minus sign in the last of these
is the difference with the CD case. When we
fold the system, the free boundary condition at $x=-L$
becomes a fixed boundary condition for the second copy. 
Thus no extra minus sign is needed to ensure fixed
boundary conditions  at $x=L$ in both copies.
Consequently, when there is no fermion backscattering,
we do not need the branch cut which we put
at the point contact at the transmitting point on the CD line.
This amounts to modifying the transmitted boundary
condition (\ref{bctransmit}) to Neumann-transmitting or `N-transmitting':
\begin{equation*}
\hbox{N-transmitting: }\:\:
\psi_{1L}(0)=\psi_{2R}(0),\:\: \psi_{1R}(0)=\psi_{2L}(0)\ .
\label{bcpinned}
\end{equation*}
so that the reflection matrix is
$${\cal R}_{\rm N-transmit} = \sigma^x\ ,$$ as opposed to
$-i\sigma_y$ for transmitting boundary conditions in the absence of
the pinned vortex, eqn.\ (\ref{Rtransmit}).

Fermion backscattering is, again, a marginal perturbation:
\begin{equation}
{L}_{fv} = i\lambda_{fv}(\psi_{1L}(0)\psi_{1R}(0) -
\psi_{2L}(0)\psi_{2R}(0))\ .
\label{Lfv}
\end{equation}
The solution can, once again, be expressed
most simply in terms of a phase shift for a
Dirac fermion: 
\begin{equation}
{\Psi_R}(0) = e^{i\widetilde{\delta}}\,{\Psi_L}(0)\ ,
\end{equation}
where $\tan(\widetilde{\delta}/2) = \lambda_{fv}/2{v_n}$.
However, we must define $\Psi$
a little differently as a result of the absence of a minus
sign in the definition of $\psi_{2L}$: 
${\Psi_R}=\psi_{2R}+i\psi_{1R}$ and ${\Psi_L}=\psi_{1L}+i\psi_{2L}$.
Consequently, the reflection matrix now takes the form:
\begin{eqnarray}
\label{R-general-N}
{\cal R}(\lambda_{fv}) &=&
\cos\widetilde{\delta} \, {\sigma^x} + \sin\widetilde{\delta} \, {\sigma^z}
\cr
 &=& \frac{1-({\lambda_{fv}}/2)^2}{1+({\lambda_{fv}}/2)^2}
\, {\sigma^x} 
+ \frac{\lambda_{fv}}{1+({\lambda_{fv}}/2)^2}\, {\sigma^z}
\end{eqnarray}

To match the results with those from bosonization, we form the Dirac
fermion $\chi$ precisely as in the last subsection:
${\chi_R}=-\psi_{1R}+i\psi_{2R}$, ${\chi_L}=\psi_{1L}+i\psi_{2L}$.
The CN boundary condition is now ${\chi_R}(0) = i e^{i\widetilde{\delta}}
{\chi_L}(0)$.  Bosonizing as before, we see that the dual boson
$\widetilde{\varphi}=\frac{1}{2}(\varphi_R-\varphi_L)$ now has CN
boundary condition $2\widetilde{\varphi}_0\equiv
2\widetilde{\varphi}(0)=\widetilde{\delta}+\pi/2$.  Once again, we have a critical
line parametrized equivalently by $\lambda_{fv}$, $\widetilde{\delta}$, or
$\widetilde{\varphi}_0$.  

When there is no backscattering we have the Neumann-transmitting
boundary condition, $\widetilde{\varphi}_0=\pi/4$.
For $\widetilde{\delta}=\pi/2$ or $\widetilde{\varphi}_0=\pi/2$ on the other hand,
every Majorana fermion incident on the point contact is backscattered:
\begin{equation}
\hbox{(free,$\pm$): } 
\psi_{1R}(0)=\psi_{1L}(0),\quad \psi_{2R}(0)=-\psi_{2L}(0)\ .
\label{free-pm}
\end{equation}
Thus the droplet is broken in two. From (\ref{R-general-N}),
we see that one droplet has free boundary condition
while the other has fixed boundary condition, but with
either fixed value equally likely. Like the $\widetilde{\delta}=-\pi/2$ case on
the CD line, this is not a product boundary condition. In analogy with
the notation, we call this (free,$\pm$).
For $\widetilde{\delta}=-\pi/2$ or $\widetilde{\varphi}_0=0$,
this is reversed, and we obtain ($\pm$, free)
boundary conditions.

\subsection{The Moore-Read Pfaffian state}

Thus far, we have focused on the Majorana fermion edge mode,
ignoring the charged mode which is present
in the Moore-Read Pfaffian state
\cite{Moore91}.  The edge theory of the Moore-Read state involves a
chiral boson $\phi_c$ as well as the Ising fields $I,\sigma$ and
$\psi\,$ \cite{Milovanovic96}.  At filling fraction $\nu=1/m$, the
``electron'' on the edge is created by the operator $\psi e^{ i
\sqrt{m}\phi_c}$. Setting $m=2$ gives the case which may be relevant to
the $\nu=5/2$ quantum Hall state, so that the ``electron'' is the
physical electron. The topological field
theory is defined as the full theory ``mod an electron''; fusing with
an electron leaves the same topological state.
Each different type of quasiparticle corresponds to a primary field in a
rational conformal field theory:  at $m=2$
they comprise the Neveu-Schwarz sector of
the superconformal theory with central charge $c=3/2$, and are the
identity, $\sigma e^{\pm i\phi_c/2\sqrt{2}}$, $\psi$, and $e^{\pm i
\phi_c/\sqrt{2}}$. (If one instead considers the $m=1$ bosonic Moore-Read
state, one obtains the $SU(2)_2$ conformal field theory, which has
primaries $\sigma e^{i\phi_c/2}$ and $\psi$.)

The boundary conditions of a squashed MR state are therefore
combinations of the three boundary conditions of the critical $2D$
Ising model with those of the charge boson. The key condition is that
all allowed boundary conditions should be left invariant by adding an
electron, since an electron carries no topological charge.  The chiral
boson here has $m$ different primary fields and, therefore, $m$
different conformal boundary conditions, corresponding to the
different possible electrical charges at the edge, modulo the charge
of the electron.  Therefore, the allowed (Ising, Charge) boundary
conditions are $(+,n)+(-,n+m)$, $(-,p)+(+,p+m)$, $(f,q/2)+(f,q/2+m)$
where $n,p=0,1,\ldots,m-1$ and $q=1,3,5,\ldots,2m-1$. All of the $2m$
boundary conditions $(+,n)+(-,n+m)$, $(-,p)+(+,p+m)$ are analogous to
the fixed boundary conditions, and they can result at the boundary of
either of the two droplets formed when a droplet is broken in two by
vortex backscattering. The key difference is that electron tunneling is an
allowed (although irrelevant) perturbation, whereas only fermion
bilinears can tunnel from fixed boundary conditions in the Ising
model.  Similarly, $(f,q/2)+(f,q/2+m)$ is analogous to the free
boundary condition.

A similar analysis applies to the anti-Pfaffian state \cite{Lee07,Levin07}.
At the edge of this state, the Majorana fermion mode $\psi$ propagates in the
opposite direction to the charged mode $\phi_c$ and there
is an extra counter-propagating neutral bosonic mode $\phi_n$.
The latter can be fermionized, thereby forming, together with $\psi$,
a triplet of Majorana fermions. Consequently, there is a triplet of electron
operators, and all boundary conditions must be invariant under
adding any of these three electron operators.

Because the fermion $\psi$ has no charge, our earlier analysis of
zero-mode tunneling and fermion backscattering goes through without
modification. However, vortex backscattering is no longer the
single-channel Kondo problem discussed in the next section, but
instead becomes a variant of the two-channel Kondo problem \cite{Fendley07a}.

\section{Flows for a Majorana fermion edge mode}
\label{sec:flows}

We have described in depth in the previous sections the different
boundary fixed points possible for a Majorana fermion edge mode.
In this section we describe the flows between these fixed point
occurring when interactions at the point contact break scale and
conformal invariance.  A valuable tool in understand these flows is
``boundary'' entropy, a subleading term in the entropy which depends
on the boundary conditions \cite{Affleck91}. At boundary fixed points,
it can be computed directly from conformal field theory
\cite{Cardy89}. Moreover, it must
decrease (at least in perturbation theory) in these flows, so the
resulting constraints allowing us here to understand
essentially all the different types of flows. Some of these flows
have been understood in the Ising context
\cite{Oshikawa97,leclair99}, but new insight is gained by considering
the topological context.

%(When bulk-edge coupling is allowed, the system
%flows to a point on the fixed line discussed in the previous subsection.)
%Before studying this case, it is useful to step back
%momentarily and 
%
%considering various
%types of tunneling, and also to obtain easily the entropy drop in all these
%flows. 
%
%When vortices tunnel, (1) there is
%a crossover scale in the flow from transmitting boundary
%conditions, and (2) the flow is to (fixed,fixed) boundary
%conditions. 

\subsection{Flowing from CN to CD by zero-mode tunneling}

We have already discussed one flow between different boundary
conditions.  In section \ref{sec:free-fixed-flow} the coupling of the
edge to a bulk vortex causes a flow from free to fixed boundary
conditions, with resulting entropy change $\Delta S=-\ln\sqrt{2}$. The
obvious generalization to the point-contact case is in agreement with
the change from (free, fixed) to (fixed, fixed) in the table, which is
the flow for bulk-edge coupling in one of two decoupled
droplets. Likewise, the flow from CD to (free, fixed) must have the
same entropy drop, since (free, free) is a point on the Dirichlet
line.

As noted above, the entropy along the CN
line is higher than along the CD line by
$\ln\sqrt{2}$. When the vortex is coupled to the edge of the system
via resonant Majorana fermion tunneling, the boundary
condition flows from a point on the CN
line to a point on the CD line. The value of $\delta$ in the latter is
not $\widetilde{\delta}$ in the former, unless the vortex is coupled
to only one side of the point contact.

To explore this further, we now couple the zero mode of the pinned
vortex to the edge. The interesting difference with the calculation in
section \ref{sec:free-fixed-flow} is that we can couple the vortex
to the bottom and top edges of the point contact, with couplings
$\lambda_T$ and $\lambda_B$ respectively. The Lagrangian in folded
language is
\begin{eqnarray}
\nonumber
L_{pv} &=& i \psi_{0} \partial_t \psi_{0} + 
i \lambda_T \psi_{0} [\psi_{1R}(x=0) + \psi_{2L}(x=0)]\\&&
+i \lambda_B \psi_0 [\psi_{1L}(x=0)  + \psi_{2R}(x=0)]\ .
\end{eqnarray}
Note that because $\psi_0^2=1$, when
$L_{pv}$ is present in the Lagrangian, non-resonant
fermion backscattering (\ref{Lfv}) is generated
at order $\lambda_T\lambda_B$, i.e.\ a Majorana fermion can
tunnel to the vortex and then from there to the other side.

For simplicity, let us suppose that we are at the N-transmitting
point on the CN line.
If we were to set either $\lambda_T$ or $\lambda_B=0$,
then $L_{pv}$ would reduce precisely to (\ref{Lresonant}).
As we saw in section \ref{sec:free-fixed-flow}, the bulk
vortex is effectively absorbed by the edge to which it is
coupled, with an entropy drop of $\ln\sqrt{2}$.
The resulting boundary condition at the end of the flow is
thus the transmitting point on the CD line.
The entropy drop indeed is the difference between Neumann and Dirichlet
entropies, as seen in table \ref{tab:OA}.

For non-zero $\lambda_T\lambda_B$, the resulting flow is from
Neumann-transmitting to a point on the Dirichlet line with
$\delta\ne\widetilde{\delta}$.
Since the action remains quadratic,
we can diagonalize it explicitly to compute the reflection matrix for
any $\lambda_T$ and $\lambda_B$. In the DC limit (or equivalently,
large $\lambda_T$ and/or $\lambda_B$), we find
\begin{equation}
{\cal R}_{pv}(\theta)= \sin(2\theta) \mathds{1}  + \cos(2\theta) i\sigma^y ,
\end{equation}
where $\theta = \arctan(\lambda_T/\lambda_B)$.
In the resonant case, $\lambda_T=\pm \lambda_B$, we have $\theta=\pm
\pi/4$ so that
\begin{equation}
{\cal R}_{res} = \pm \mathds{1}  ,
\end{equation}
corresponding to complete backscattering. These are the
(free, free) and $(\pm,\pm)$ boundary conditions discussed above.
For arbitrary $\lambda_T$, $\lambda_B$, we end up at the point on
the CD line with $\delta=2\theta$ or, equivalently,
${\varphi_0}=\theta+\pi/4$. If we start at a point on the CN
line with $\widetilde{\delta}\neq 0$, then we end up at the point
$\delta=\widetilde{\delta}+2\theta$ on the Dirichlet line since the
phase shifts add.

\begin{figure}
\centerline{\includegraphics[width=85mm]{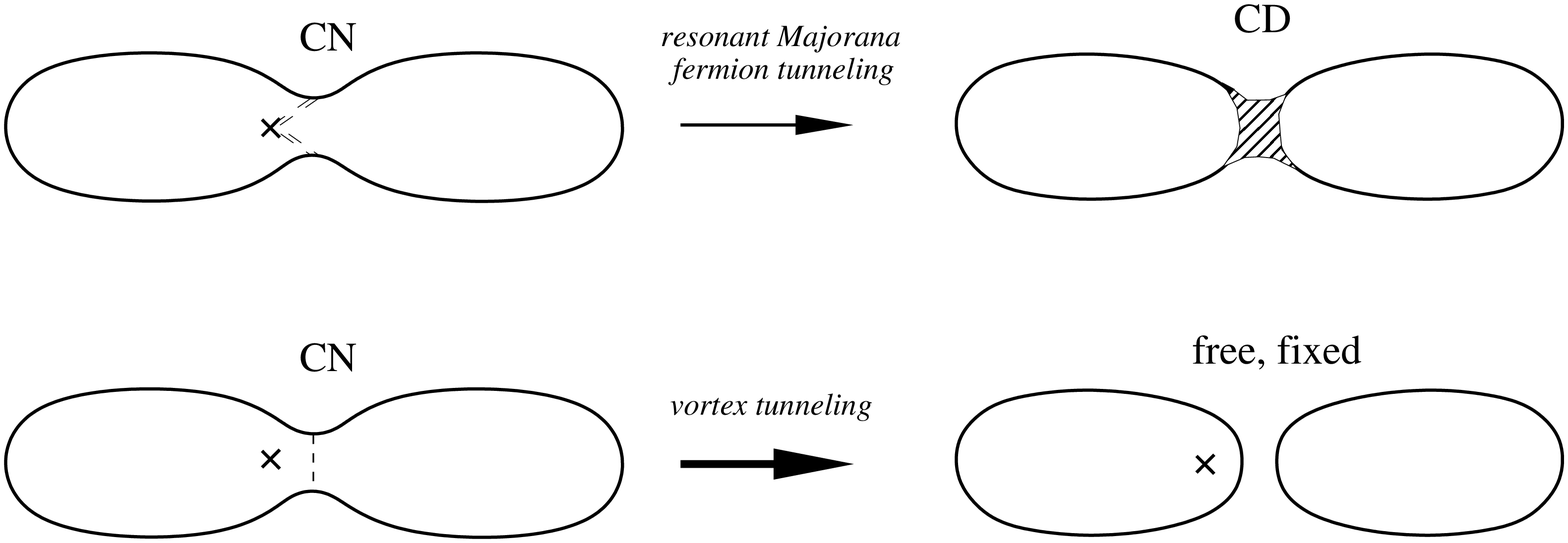}}
\caption{(a) Inter-edge {\it Majorana fermion} backscattering
through the zero mode on a localized quasiparticle
takes the system from a point on
the continuous Neumann line to a point on the continuous
Dirichlet line. The region between the two droplets is not vacuum
since inter-droplet Majorana fermion tunneling is an allowed
perturbation.
(b) Inter-edge {\it vortex} backscattering takes the system from a
point on the continuous Neumann line to the (free, fixed) point.
There is vacuum between the two droplets into which the
system is broken.}
\label{fig:pt-contact}
\end{figure}

\subsection{Vortex backscattering}
\label{subsec:Fixed-fixed}

In the previous section, we have analyzed the continuous Dirichlet and
Neumann boundary conditions in terms of Majorana fermion
backscattering at a point contact. In order to discuss the remaining
boundary conditions, (free, fixed) and (fixed, fixed), and the flows
to them from CD and CN boundary conditions, we need to recall a few
facts about inter-edge vortex backscattering.

The problem of vortex backscattering across a point contact is not
as easily solvable as fermion backscattering is. In two
earlier papers \cite{Fendley06,Fendley07a}, however, we exploited
bosonization to map this problem onto the anisotropic single-channel
Kondo problem. The effective spin-1/2 Kondo spin $\vec{S}$ arises to take
account of the non-Abelian statistics of the vortex operators. 
The vortex backscattering Lagrangian (in the normalization used above) is 
%a vortex
%tunneling between the right and left moving edges with a schematic
%Lagrangian of the form, ${\cal L}_{tun} \sim \lambda_v \sigma_R
%\sigma_L$ where $\sigma_{R/L}$ are edge vortex creation operators.  
\begin{equation}
{L}_{v} = {\lambda_v} ( S^+ e^{i\varphi(0)/2\sqrt{2}} + S^-
e^{-i\varphi(0)/2\sqrt{2}})\ .
\label{kondo}
\end{equation} 
Since the dimension of $e^{\pm i\varphi(0)/2\sqrt{2}}$ is 1/8, vortex
backscattering is relevant; it causes the system to flow away from the
Dirichlet line.  (Where we start on the Dirichlet line depends on the
strength of fermion backscattering.) Independent of where we start,
this is a strongly relevant perturbation to the free Majorana edge
modes and there is a crossover to strong coupling where all edge modes
are completely backscattered and the system breaks into two
droplets. The resulting entropy drop is $\ln 2$, due to the screening
of the emergent spin-1/2 degree of freedom. We conclude that, in Ising
language, the flow for vortex backscattering is from Dirichlet to
(fixed, fixed).  This is the most stable boundary condition: vortex
backscattering splits the droplet in two so completely that all
allowed perturbations coupling the two droplets are irrelevant
\cite{Fendley06,Fendley07a}. Thus, in the language of defect lines in
the Ising model, vortex backscattering is a magnetic field at the two
columns of sites on either side of a particular column of bonds, which
leads to fixed boundary conditions for both columns of spins. Indeed,
one can see to all orders in perturbation theory that the Kondo
interaction (\ref{kondo}) is equivalent to adding a defect magnetic
field in the Ising field theory \cite{Lamacraft08}.  In contrast, Majorana
fermion backscattering is, as we have seen, a weakening (or
strengthening) of that column of bonds.

By the same logic that led us to conclude that
vortex backscattering leads from the CD line to
the (fixed, fixed) point, we conclude that when there is a single
vortex in the bulk, decoupled from the edge, vortex backscattering
causes the flow from CN to (free, fixed)
boundary conditions. Of course, unless the bulk vortex is
right at the point contact, this is really the same, as far as local
physics near the point contact is concerned, as the Dirichlet
to (fixed, fixed) flow since the branch cut associated with the
bulk vortex can be moved to $x=L$ by a gauge transformation.
When the bulk vortex is right at the point contact, however,
the problem is quite subtle and depends on the precise backscattering
paths and how they wind around the bulk vortex.

A simple picture explains the difference
between the (free, free) and (fixed, fixed) points
heuristically. At both fixed points, the system divides
into two droplets. Imagine tunneling a fermion from
one droplet to the other; the tunneling term would
look like:
\begin{equation}
H_{\rm tun} = {t_f} \left(\psi_{1R}(0)+\psi_{1L}(0)\right)
\left(\psi_{2R}(0)+\psi_{2L}(0)\right)
\end{equation}
In the (free, free) case, $\psi_{1R}(0)=\psi_{1L}(0)$
and $\psi_{2R}(0)=\psi_{2L}(0)$. Hence, it is possible
to couple the two droplets with such a term. It
is simply that ${t_f}$ has been tuned to zero at the
(free, free) point. By varying ${t_f}$, we move along the
CD line. In the (fixed, fixed) case,
however, $\psi_{1R}(0)=-\psi_{1L}(0)$
and $\psi_{2R}(0)=-\psi_{2L}(0)$. Therefore,
$H_{\rm tun}$ vanishes at the fixed point, and single
fermion tunneling (which would be a marginal perturbation)
is not possible. Instead,
the leading perturbation of the fixed point is
a Cooper pair tunneling term \cite{Fendley06,Fendley07a}.
The same argument applies to the (free, fixed) point.
Meanwhile, the $(\pm,\pm)$ point and the continuous
Neumann analogues of (free, free) and $(\pm,\pm)$
can all be perturbed by fermion tunneling.

\subsection{Entropy drops}

\begin{figure}
\centerline{\includegraphics[width=75mm]{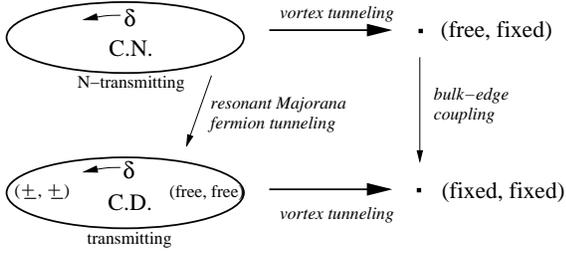}}
\caption{The different fixed lines and fixed points
for a point contact in the Majorana fermion edge mode. The entropy drop
associated with resonant Majorana fermion
backscattering (CN to CD and (free, fixed) to (fixed, fixed))
is $\ln\sqrt{2}$. The entropy drop
associated with vortex backscattering (CN to (free, fixed)
and CD to (fixed, fixed)) is $\ln 2$.}
\label{fig:contact-pd}
\end{figure}

We have shown in this section that there are a variety of flows
between the various boundary fixed points.
We summarize these flows in
the figure \ref{fig:contact-pd}. In this subsection we discuss these
flows in more depth by analyzing the entropy drops in these flows.

In Refs.\ \onlinecite{Fendley06,Fendley07b}, it was noted that the
entropy drop which takes place as a result of vortex backscattering
(i.e.\ as the system flows from the Dirichlet line to (fixed, fixed)
or from the Neumann line to (free, fixed)) could be understood simply
as the result of one droplet breaking into two.  Namely, the
subleading term in the thermodynamic entropy that we have discussed
here has an intriguing correspondence with a different object, the
entanglement entropy between regions in a topological state
\cite{Kitaev05,Levin05}.  The topological entanglement entropy of a
droplet with trivial total topological charge and perimeter $L$ is
$S=\alpha L - \ln{\cal D}$, where $\alpha$ is temperature-dependent
and ${\cal D}$ is the total quantum dimension of the particular
topological state of matter.  For a $p+ip$ superconductor, ${\cal
D}=2$. When a droplet breaks into two droplets, each of which has
trivial topological charge, the entropy of the two droplets is
$S=\alpha{L_1} + \alpha{L_2} - 2\ln{\cal D}$. Hence, if the total
length of the edge(s) remains unchanged, $L={L_1} + {L_2}$, the
entropy drop is simply $\ln{\cal D}=\ln 2$. This $\ln 2$ entropy drop
as is the same as the drop in the thermodynamic entropy resulting from
vortex backscattering across a point contact effectively splitting the
droplet in two. This correspondence is not a coincidence, but has been
proved to hold for arbitrary topological states \cite{Fendley07b}.

In light of this correspondence, it is natural to wonder why there is
no entropy drop associated with breaking the system in two by tuning
from the transmitting point $\lambda_f=0$ to ${\lambda_f}=\pm 2{v_n}$,
i.e.\ the (free, free) and $(\pm,\pm)$ points. The difference with the
(fixed, fixed) case, which does have an entropy drop, is seen most
easily at the $(\pm,\pm)$ point. At the $(\pm,\pm)$ point, the system
is broken in two, just as in the (fixed, fixed) case, so there is a
naive entropy drop of $\ln 2$. However, the spins at the defect line
are equally likely to both be fixed up as to be fixed down.  Hence,
there is an {\em extra} $+\ln 2$ entropy associated with the choice of
up or down spin.  This cancels the naive entropy drop, so the total
entropy is unchanged.  To what does this choice of up or down spin
correspond in fermionic language? Let us suppose that the fixed
boundary conditions at $x=L$ are up (i.e.\ $+$) in both copies, so
that the total fermion number is even (as discussed in sections
\ref{sec:Majorana-Ising} and \ref{sec:free-fixed-flow}). Then, if the
spins at $x=0$ are both up, the fermion number in both droplets is
even. If the spins at $x=0$ are both down, then both droplets have odd
fermion number. Both possibilities are equally likely at the
$(\pm,\pm)$ point. In effect, Majorana fermion backscattering, when
tuned to $\delta=-\pi/2$, generates a qubit, just as if we had a pair
of vortices in each half of the original droplet (with the total
topological charge of all four vortices being trivial). Since the
fermion number parity is the same in each droplet, the two droplets
are not really decoupled.  However, vortex backscattering causes the
fermion number parity in each droplet to be fixed; the $\ln 2$ entropy
drop can be interpreted as the loss in the uncertainty in the fermion
number parity as the two droplets become truly decoupled.

At the (free, free) point, each droplet has a zero mode.
Let us call the corresponding operators $\psi_0^L$, $\psi_0^R$,
%Then, if the total fermion number is even, the system
%must be in an eigenstate of $i{\psi_0^L}{\psi_0^R}$.
%(Whether the eigenvalue is $-1$ or $+1$ is a matter of
%a gauge choice.)
and consider the operator $(-1)^{N_F^R}$,
the parity of the fermion number on the right droplet.
This operator satisfies the anti-commutation relation
$\{(-1)^{N_F^R},i{\psi_0^L}{\psi_0^R}\}=0$.
Representing this algebra requires a two-dimensional ground
state Hilbert space \cite{Ginsparg}.
Hence, there is, once again, a $+\ln 2$ entropy. However,
unlike in the $(\pm,\pm)$ case, this entropy can be
split into a $\ln\sqrt{2}$ entropy ascribed to each droplet.
When vortex backscattering is turned on, it is as if a magnetic field
has been applied to the free boundaries of both edges:
they both flow to fixed, losing entropy $2\ln\sqrt{2}$.

On the CN line, we similarly have a naive
entropy drop of $\ln 2$ when the droplet is split into two
at the (free,$\pm$) and ($\pm$, free) points. However,
there is, once again, an uncertainty in the fermion number
parity of each droplet. One of the droplets contains the vortex
and, therefore, has a zero mode. The other droplet can have either
even or odd fermion number, as if it contained a pair of vortices,
and this uncertainty leads to an extra entropy of $\ln 2$.

The CN to CD flow is
caused by the coupling of a bulk vortex to the edge,
which causes an entropy drop of $\ln\sqrt{2}$.
The flow from CD to (fixed, fixed) is caused
by inter-edge vortex backscattering, as is the flow from
CN to (free, fixed); both lead to an entropy
drop of $\ln 2$. The flow from (free, fixed)
to (fixed, fixed) is again caused by the coupling of a bulk vortex
in the left droplet to the edge, accompanied by an entropy
drop of $\ln\sqrt{2}$. The only remaining flow is from
CD to (free, fixed), which must also be
accompanied by an entropy drop of $\ln\sqrt{2}$.
At the (free, free) point on the CD line,
it is simply the flow of the $x=0$ boundary of one of the droplets
from free to fixed boundary conditions. Since the droplets
have zero modes, this flow is presumably caused by the coupling
of this zero mode to the edge.

The following is one way to view the results of this section.
When vortex backscattering is allowed, the droplet breaks
completely in two and there is vacuum between the two
droplets. Thus, as suggested in Refs.\ \onlinecite{Fendley06,Fendley07b},
the entropy drops by $\ln 2$ since the number of droplets
increases by one (or the Euler characteristic increases by
one, as alluded to above). However, when only Majorana
fermion backscattering is allowed, even when the system is
tuned to the (free, free) point and the droplets seemingly
break in two, it isn't vacuum between the droplets.
Instead, another topological phase is effectively nucleated
which has Majorana fermions as an allowed (though gapped)
excitation. The obvious choice is the toric code (the doubled
$\mathbb{Z}_2$ phase), which occurs in the bulk when
the Chern number of the Majorana fermion number excitations
of the Ising phase changes \cite{Kitaev05a} (see also
Ref. \onlinecite{Bais08}). Since this phase has
the same total quantum dimension ${\cal D}$, it is natural
that there be no entropy drop when it is nucleated.

We conclude this section by noting that these arguments about entropy
drops hold in arbitrary geometries. In particular, the correspondence
with topological entanglement entropy still holds.  As shown by
Bonderson\cite{Bonderson09}, the entanglement entropy of a topological
fluid in any planar region $R$ is equal to
\begin{equation}
S = \alpha L - {\chi^{}_R}\ln{\cal D}
\end{equation}
where $\chi^{}_R$ is the Euler characteristic of the region $R$ and
${\cal D}$ is the total quantum dimension of the topological
fluid. Splitting a disc into two causes $\chi_R$ to increase from $1$
to $2$, so indeed the entropy drop is $\ln{\cal D}$.  A more
complicated situation is if the topological fluid has an annular
topology. As discussed in section \ref{sec:point-contact}, the annulus
corresponds to choosing transmitting boundary conditions at $x=L$. In
the $p+ip$ case, the classical analog is an Ising model with periodic
boundary conditions and a defect line. At the point contact at which
vortex backscattering is allowed, the flow effectively splits the
system at the point contact. The presence of the transmitting boundary
conditions at the other end does not change this, as makes sense
physically. However, now splitting the system at the point contact
does not split the system into two, but rather changes the annular
geometry to that of a single disc, changing the Euler characteristic
from 0 to 1. Thus even though the entropies are different due to the
different geometries, the entropy drop in the flow between the remains
the same $\ln{\cal D}$. In fact, whenever any geometry is split via a
point contact, the Euler characteristic will always decrease by $1$,
in accord with our assertion that the dynamics of the point contact is
not affected by the geometry of the full system (i.e.\ the other
boundary conditions).

\section{Multiple bulk quasiparticles}
\label{sec:many-qps}

In the foregoing we have dealt exhaustively with the cases of no
vortices and a single vortex in the bulk of a $p+ip$
superconductor. We saw how in the absence of tunneling and
backscattering, the effect of a bulk vortex could be taken into
account by changing a boundary condition in the squashed
system. Zero-mode tunneling between this vortex and the edge then
causes a change in this boundary condition. We devote this section to
explaining how to account for the effect of multiple bulk
quasiparticles on the edge modes.  This makes it possible to consider
different types of tunneling processes, including zero-mode tunneling
between vortices.

In this section we again focus on a Majorana fermion edge mode, but it is
straightforward to generalize the results: one of our earlier
papers \cite{Fendley07b} contains most of the necessary
analysis. There we showed how to compute the partition function of the
chiral conformal field theory describing the edge modes of a
topological fluid. This partition function depends on how many of each
type of quasiparticle are in the bulk.  We dubbed this the
``holographic'' partition function, because it encodes the topological
properties of the full system, not just the edge. For example, the
universal part of the full topological entanglement entropy can be
extracted from these partition functions. Squashing the system of
course does not change the partition function, so our earlier
computations still apply.

\begin{figure}
\centerline{\includegraphics[width=65mm]{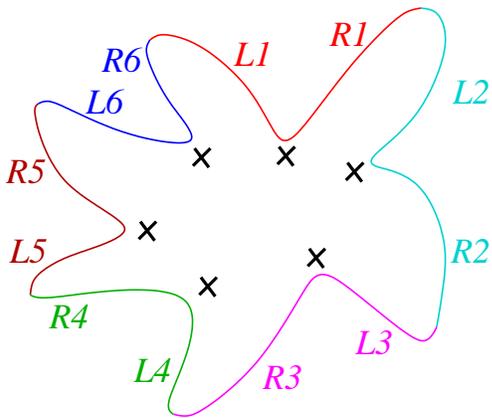}}
\caption{A droplet with 6 quasiparticles, each marked by a cross. This
  can be mapped to 6 non-chiral systems on the strip, so the each line
  segment labeled by $Ri$ or $Li$ becomes the right or left movers in
  the $i$th copy. The successive copies are coupled at the ends of the
  ``fingers'', while the left and right movers within the $i$th copy
  are coupled at the point on the edge closest to the $i$th vortex.
 For a Majorana fermion edge mode, these boundary conditions are given
  explicitly in (\ref{atvortex},\ref{fingerend}).}
\label{fig:fingers}
\end{figure}

As before, the goal is to squash the droplet with chiral edge modes
down to a line segment having both left and right movers. However, in
order to be able to treat the many different possible tunneling
process in terms of (perturbed) boundary conformal field theory, we
must make multiple copies of the system. There is not a unique way of
doing this, but one convenient way of doing so is illustrated in
figure \ref{fig:fingers}. When there are $n$ quasiparticles, then we
have $n$ copies of the system. The $x=0$ boundary of each of the $n$th
copy corresponds to the point on the edge closest to the $n$th
quasiparticle. The boundary condition on each of these copies then
depends on the type of the corresponding quasiparticle. For a vortex
in the $p+ip$ case, this is of course the free boundary condition.
The $x=L$ boundary is where we couple the copies with a
``fixed-transmitting'' boundary condition. This means that we couple
the left movers in the $i$th copy to the right movers in the $i+1$st
copy (mod $n$) at $x=L$. 

For the $p+ip$ superconductor with $n$
vortices, putting this together means
\begin{eqnarray}
\label{atvortex}
\psi_{iL}(0)&=&\psi_{iR}(0),\\
\label{fingerend}
\psi_{iR}(L)&=&(-1)^{i+1}\psi_{(i+1)L}(L)\ .
\end{eqnarray}
for all $i$. The effect of zero-mode tunneling here between the $i$th
vortex and the edge then results simply in adding to the Hamiltonian
the term (\ref{Lresonant}) for the $i$th copy.

For one vortex, these boundary conditions reduce to those considered
previously. For two vortices, we can simplify the problem somewhat. As
discussed at the end of section \ref{sec:Majorana-Ising}, one doesn't
need two copies here: one can impose free boundary conditions at both
ends of a single copy. This can be recovered from the above by
treating the boundary condition (\ref{fingerend}) at $x=L$ for $n=2$
as transmitting. Then unfolding indeed gives a system with free
boundary conditions at both ends of a system of length $2L$. The
advantage of using two copies instead of one is that then one can
consider tunneling between the two vortices as a boundary condition,
in the same framework as all our calculations: it is
a dimension-0 perturbation which is, therefore, strongly
relevant and causes a flow to the state with trivial total
topological charge and entropy diminished by $\ln 2$.

Including a point contact is not much more difficult.  One just needs
two of the ``inside'' points in figure \ref{fig:fingers}, and the point
contact couples the two systems at $x=0$, in the fashion as
before. The system can then be split into two across this point
contact, just as before. Similarly, a hole can be put in the surface
by changing the boundary conditions at $x=L$ so that e.g.\
$\psi_{iR}(L)=\psi_{(i)L}(L)$ and $\psi_{(i-1)R}(L)=\psi_{(i+1)L}(L)$,
just like we did for the annulus.

\section{Fibonacci Anyons and the 3-State Potts
Model}
\label{sec:potts}

It has long been known that each primary field in a rational conformal
field theory corresponds to a particular conformal boundary condition
\cite{Cardy89}. In the context of a topological state, this means that
there is a boundary fixed point corresponding to each quasiparticle
type. This correspondence has a very nice application in this context:
it means we can determine how boundary conditions change when
additional bulk quasiparticles are included. This is
the topological analog of fusion by a boundary operator
\cite{Cardy89,Affleck91,Fuchs01}; the corresponding chiral 
partition functions can be found in ref.\ \onlinecite{Fendley07b}.

We now illustrate this in a somewhat more complicated example
of a topological state. The Read-Rezayi (RR) states \cite{Read99}
generalize the MR state by replacing the
($k$=$2$) Ising CFT with the $\mathbb{Z}_k$ parafermionic model,
while simultaneously changing the radius
of the charge boson in order to keep the scaling dimension
of the electron operator fixed. In this section,
we will briefly consider the first state after the
Moore-Read state in this sequence, the $k=3$
Read-Rezayi state. If a quantum Hall state were observed
at $\nu=13/5$, this state might be realized there.
The particle-hole conjugate of the Read-Rezayi state,
the anti-RR state \cite{Bishara08}, might
be realized at the observed plateau at $\nu=12/5$.
The anti-RR state is related to the RR state
in the same way as the anti-Pfaffian state is related to
the MR state. Thus, much of what we have to say
about the $k=3$ RR state applies as well to the $k=3$
anti-RR state.

The neutral sector of the edge theory of the $k=3$ RR state is the
$\mathbb{Z}_3$ parafermionic CFT.  This conformal field theory
describes the critical ferromagnetic $3$-state Potts model. The
$3$-state Potts model is a lattice model of spins $s_i$ which take the
values ${s_i}=A, B, C$. The Hamiltonian is
\begin{equation}
H = -J\sum_{\langle i,j\rangle} \delta_{{s_i}{s_j}}
\end{equation}
This model undergoes a second-order phase transition
at a temperature $T_c$ below which the spins develop
a non-zero expectation value.
The conformal theory for the critical point has 6 primary
fields, $1$, $\psi$, $\psi^\dagger$, $\sigma$, $\sigma^\dagger$,
$\varepsilon$. The fields $1$, $\psi$, $\psi^\dagger$
have a $\mathbb{Z}_3$ structure to their fusion rules:
$\psi\times\psi=\psi^\dagger$, $\psi^\dagger\times\psi^\dagger=\psi$,
$\psi\times\psi^\dagger=1$.
The field $\varepsilon$ is a Fibonacci anyon:
$\varepsilon\times\varepsilon=1+\varepsilon$.
Then $\sigma$, $\sigma^\dagger$ are formed
by combining $\varepsilon$ with the $\mathbb{Z}_3$ fields:
$\sigma=\varepsilon\times\psi$,
$\sigma^\dagger=\varepsilon\times\psi^\dagger$.

In the $k=3$ RR quantum Hall state,
there is also a charge boson, and the quasiparticles
of the theory are products of the
critical Potts quasiparticles with exponentials of the charge boson.
This state can be interpreted as a quantum Hall state of
triplets of electrons. The parafermions $\psi$, $\psi^\dagger$
can then be viewed as a single electron or
a pair of electrons (modulo $3$). The Potts spin operator
$\sigma$ can be viewed as a flux $hc/3e$ vortex,
and $\sigma^\dagger$, $\varepsilon$ can be viewed as such 
a vortex with one or two parafermions in its core, respectively.
When the electrical charge is included, $\sigma$ has electrical
charge $1/5$, as we would expect for a flux $hc/3e$ vortex
at $\nu=3/5$; $\sigma^\dagger$ has electrical charge
$3/5$, as we would expect for a vortex with a charge $2/5$
parafermion; $\varepsilon$ is neutral, as we would expect
for a vortex with two parafermions.

All the conformal boundary conditions in the three-state Potts model
are known \cite{Affleck98}.  There are six boundary conditions
corresponding to the six quasiparticles (ignoring extra multiplicities
due to the different possible electric charges). These are the three
possible fixed boundary conditions, denoted $A$, $B$ and $C$, and
three more in which one of the spin values is forbidden at the
boundary, denoted not-$A$, not-$B$, and not-$C$ \cite{Saleur88}. With
the not-$A$ boundary condition, each boundary spin is independently
allowed to take either the value $B$ or $C$ while the value $A$ is
forbidden. These boundary conditions are called mixed
boundary conditions, and sometimes not-$A$ is written $BC$.

Consider a droplet which has been squashed, as in our
Ising discussion, and suppose that the boundary condition
at $x=L$ is fixed to $A$. When there are no quasiparticles
in the bulk, the boundary condition at $x=0$ will also
be fixed to $A$. If there is a $\psi$ or $\psi^\dagger$
in the bulk, it will, instead, be fixed to $B$ or $C$, respectively.
(In the quantum Hall context,
a $\psi$ must be accompanied by an electric charge
and a corresponding boson exponential in order to
be local with respect to electrons. However, the charge
part of this quasiparticle plays no role in the present discussion.)
All three of these possible fixed boundary conditions are
completely stable (assuming that the bulk quasiparticle
is pinned). Since $\psi$ and $\psi^\dagger$ are Abelian
quasiparticles, this is not surprising.
If there is an $\varepsilon$ in the bulk,
the boundary condition at $x=L$ will be not-$A$.
If there is a $\sigma$ or $\sigma^\dagger$ in the bulk,
the boundary condition will be not-$B$ or not-$C$, respectively.
These boundary conditions have a higher entropy
than the fixed boundary condition
by $\ln\tau$, where $\tau=(1+\sqrt{5})/2$ is the quantum
dimension of $\sigma$, $\sigma^\dagger$, and $\varepsilon$.
Thus, mixed boundary conditions are unstable to bulk-edge
coupling. The three types of vortices have a $\varepsilon$
zero mode which couples with the $\varepsilon$ field
on the edge. This is a dimension-$2/5$ perturbation of the
edge and is, therefore, highly relevant. It leads to the vortex
being absorbed by the edge and an entropy drop $\ln\tau$.

Because the $3$-state Potts model has a non-diagonal partition
function, the six boundary conditions described above
do not exhaust the conformal boundary conditions: there are in fact
eight of them \cite{Affleck98,Behrend98}. The free boundary condition is
obviously one of the missing ones.  Because it does not correspond to
one of the quasiparticle types, it does not occur simply by adding a
quasiparticle in the bulk.  (This is clear from the preceding
discussion since we have already exhausted all of these
possibilities.) It does not result from fusion with one of the primary
fields of the Potts model but, instead, from fusion with one of the
primaries of a $c=4/5$ CFT with more primary fields (the tetracritical
Ising model) \cite{Affleck98}.

Thus, the free boundary condition and the eighth boundary condition,
simply called `new' by Affleck {\it et al.} \cite{Affleck98}, are
slight oddities in the boundary conformal field theory context.
However, in the quantum Hall
context, free boundary conditions arise quite naturally in a manner
analogous to their appearance in the analysis of a point contact in
the Ising case.  Namely, consider a $k=3$ RR droplet with a point
contact in the middle.  Let us suppose that there is non-zero
backscattering of $\epsilon$ quasiparticles but no other tunneling is
allowed so that the backscattering Hamiltonian is of the form
\begin{equation}
H_{\rm tun} = {\lambda_\epsilon} \,{\epsilon_R}(0){\epsilon_L}(0)
\label{eqn:e-backscattering}
\end{equation}
The analogous Ising operator,
${\psi_R}(0){\psi_L}(0)$, has scaling dimension $1$ and, therefore, is
an exactly marginal perturbation. Here, this operator has scaling
dimension $4/5$, so that ${\lambda_\epsilon}$ is relevant and flows to
strong coupling.  To what does it flow in the strong coupling limit?
In the $3$-state Potts context, (\ref{eqn:e-backscattering}) is the
energy operator, so on the lattice it is a local change of $J$ on a
column of bonds (precisely as in the Ising case with
${\psi_R}{\psi_L}$). If ${\lambda_\epsilon}>0$, then this is a local
{\it decrease} of $J$, which flows to $J=0$, decoupling the Potts
model into two halves and leaving free boundary conditions on each
half at the column of vanishing bonds. The boundary entropy of
(free,free) boundary conditions is less than zero \cite{Affleck98}, so
this flow from transmitting from boundary conditions (where the
boundary entropy is zero) indeed has an entropy drop.

If, on the other hand, ${t_\epsilon}<0$, then this is a local
{\it increase} of $J$, which flows to $J=\infty$. As in
the Ising case, the system is again cleaved in two with the same fixed
boundary condition on each half and all three possible values of the
fixed boundary condition equally likely. We thus call this boundary
condition $(A/B/C,A/B/C)$. The parafermion number of each droplet is
not fixed, so any of the possible fixed boundary conditions can
occur. Since there are three possible values for the fixed boundary
condition, there is an entropy $\ln 3$ larger than that of fixed boundary
conditions on both droplets. 
Since this entropy is shared equally between the two droplets,
each has an entropy $\ln\sqrt{3}$ higher than
the fixed boundary condition.

(Free,free) boundary conditions have 
the same entropy as $(A/B/C,A/B/C)$. Although there is not as simple
an argument as in the case of the latter, the
underlying reason is the same, namely that the parafermion number is not
fixed. (We note that the interpretation of the entropy in terms of
the parafermion number, as with the fermion number in the
Ising case, makes sense at the fixed point in the limit of finite $L$.
While the fixed point can be reached by fine-tuning, it is generically reached
by flowing to the infrared, $T\rightarrow 0$. However, the entropy
is ordinarily computed in the opposite limit: $L\rightarrow\infty$
first and then $T\rightarrow 0$. Thus, our intuitive argument
applies to the opposite of the ordinary order of limits. The result is
the same, however, as may be seen from direct computation in the $L\rightarrow\infty$,
$T\rightarrow 0$ limit \cite{Affleck98}. The equality of these two orders of limits
may be related to the fact that we are discussing integer-valued degeneracies.)
The computation of the partition function
\cite{Cardy86b} shows that the free
boundary condition on a single system indeed has entropy $\ln\sqrt{3}$
relative to the fixed boundary condition \cite{Affleck98}.
Since $\sqrt{3}>(1+\sqrt{5})/2>1$, the free boundary condition
has higher entropy than either mixed or fixed and the addition of a
perturbation can lead to a flow to either one.  In both the (free,
free) and $(A/B/C,A/B/C)$ cases, parafermions can tunnel from one
droplet to the other, so it is clear that the two droplets are not
separated by vacuum. The natural guess is that they are separated by a
topological phase described by the deconfined phase of $\mathbb{Z}_3$
gauge theory, but this will be discussed elsewhere.

The `new' boundary condition does not have a simple interpretation in
the language of the $3$-state Potts model. However, the `new' boundary
condition is known to be dual to the mixed boundary condition(s), just
as the free boundary condition is dual to the fixed boundary
condition(s) \cite{Affleck98}.  Thus, just as the mixed boundary
condition is obtained from the fixed one by creating an additional
$\epsilon$ quasiparticle in the bulk (or $\sigma$ or $\sigma^\dagger$
for its $\mathbb{Z}_3$ partners), the `new' boundary condition is
obtained from the free boundary condition by creating an additional
$\epsilon$ quasiparticle in the bulk. Since the free boundary
condition is $\mathbb{Z}_3$-invariant, it does not matter whether we
create an $\epsilon$, a $\sigma$, or a $\sigma^\dagger$.
Alternatively, we could begin with an RR droplet with an $\epsilon$ in
the bulk (analogous to N-transmitting) and then turn on
(\ref{eqn:e-backscattering}). Either way, we see that the `new'
boundary condition has entropy $\ln\tau$ larger than the free boundary
condition.

One can handle other conformal boundary conditions in a point contact
in a $k=3$ RR droplet in a similar fashion. However, the number of
the possible boundary conditions increases rapidly as $k$ is increased. In Ising, there
are just three basic transmitting-type boundary conditions: the basic
one ($\delta=0$ on the CD line), the antiferromagnetic defect
($\delta=\pi$ on the CD line), and the N-transmitting one
($\widetilde\delta=0$ on the CN line). (The latter two defects can be fused
to give $\widetilde\delta=\pi$ on the CN line.) For the
three-state Potts model, even ignoring the charge mode, there are
already 16 of them \cite{Petkova2001}. Thus there will be an array of
unstable boundary fixed points here, even without considering fixed
points such as $(A/B/C,A/B/C)$, which is neither transmitting nor
a product boundary condition. However, there are no fixed lines for
$k=3$, because there are no dimension-1 tunneling operators.

\section{Discussion}
\label{sec:discussion}

The preceding discussion can be generalized
to other topological states. We can squash a
droplet as before so that we formulate the
edge effective theory in terms of conformal field
theory on a strip. The presence of a
quasiparticle in the bulk changes the boundary
condition at one end of the strip. The boundary
condition with no quasiparticles in the bulk
is stable. If the added quasiparticle
is Abelian, then the new boundary condition
has the same entropy as the no-quasiparticle
boundary condition and is also stable.
However, if the added quasiparticle is non-Abelian,
then the new boundary condition has higher entropy.
Such a boundary condition is unstable to coupling
to the edge. The non-Abelian quasiparticle has a
zero mode to which edge quasiparticles with
scaling dimension $\Delta<1$ can tunnel resonantly.

The general framework, which the Ising and $3$-state Potts models
exemplify, makes the notion of squashing precise.  For a
2+1-dimensional topological theory on a disk geometry, the
topologically-distinct excitations correspond to the primary fields of
the associated edge conformal field theory. If there is an excitation
labeled by $a$ in the bulk, then in the chiral conformal field theory
of the edge, which is defined on a cylindrical spacetime, the
partition function $\chi_a$ results \cite{Fendley07b}. The results of
Ref.\ \onlinecite{Cardy89} then allow one to find boundary conditions
on the strip which give the same partition function $\chi_a$. Thus any
quantity in a given sector of the chiral conformal field theory can be
computed in the non-chiral theory on the strip by imposing the
appropriate boundary conditions.

We discussed these ideas in detail in the context of 
the Ising model, where there are three primary fields,
$\sigma,1,\psi$, which correspond to free, fixed $+$, and fixed $-$,
respectively. In the corresponding $2+1$-dimensional topological
state, these correspond to a state with a vortex in the bulk, and
then the states in which the vortex has been absorbed by the edge,
with either $0$ or $1$ unpaired fermions in the bulk.

These ideas can be generalized to the description of
point contacts in topological states. Inter-edge quasiparticle
backscattering generically splits a droplet into two,
with one of the aforementioned conformal boundary conditions
on each of the resulting droplets. (Here, the Ising model
is non-generic because it has two fixed lines.)
One interesting feature which arose in our analysis is that
the region between the two droplets is generically not
the vacuum but, rather, a {\it different} non-trivial topological
phase. This may be a zero-dimensional analogue of the
condensation phenomena discussed in Refs.~\onlinecite{Bais08,Gils08}.

\bigskip
We would like to thank P.\ Bonderson, C.\ Kane, and I.\ Runkel for
very helpful conversations. This work has been partially supported by
the NSF under grants DMR/MSPA-0704666 (PF) and DMR-0529399 (MPAF).

%\end{document}

\end{document}